\documentclass{article}

\usepackage{amsmath}
\usepackage{amsfonts}
\usepackage{graphicx}
\usepackage{epstopdf}
\usepackage{multirow}
\usepackage{cite}
\usepackage{a4wide}
\usepackage{color}
\usepackage[Symbol]{upgreek}
\usepackage{authblk}

\definecolor{grey}{rgb}{0.66,0.66,0.66}

\def\DKL{D_\mathrm{KL}}
\renewcommand{\vec}[1]{\boldsymbol{\mathbf{#1}}}
\newcommand{\fit}[1]{\hat{#1}} 
\newcommand{\obs}[1]{\tilde{#1}} 
\newcommand{\fitp}{\fit{p}}

\newcommand{\fitvp}{\fit{\vec{p}}}
\newcommand{\fitvq}{\fit{\vec{q}}}
\newcommand{\mat}[1]{\boldsymbol{\mathbf{#1}}}
\def\mpi{\uppi}
\newcommand{\obsp}{\obs{p}}
\newcommand{\obsvp}{\obs{\vec{p}}}

\date{}
\begin{document}

\title{Cross-sectional Markov model for trend analysis of observed discrete distributions of population characteristics}
\author{Agnieszka Werpachowska\footnote{aw@averisera.uk}~\ and Roman Werpachowski}
\affil{Averisera Ltd, London, United Kingdom}

\maketitle

\begin{abstract}
We present a stochastic model of population dynamics exploiting cross-sectional data in trend analysis and forecasts for groups and cohorts of a population. While sharing the convenient features of classic Markov models, it alleviates the practical problems experienced in longitudinal studies. Based on statistical and information-theoretical analysis, we adopt maximum likelihood estimation to determine model parameters, facilitating the use of a range of model selection methods. Their application to several synthetic and empirical datasets shows that the proposed approach is robust, stable and superior to a regression-based one. We extend the basic framework to simulate ageing cohorts, processes with finite memory, distinguishing their short and long-term trends, introduce regularisation to avoid the ecological fallacy, and  generalise it to mixtures of cross-sectional and (possibly incomplete) longitudinal data. The presented model illustrations yield new and interesting results, such as an implied common driving factor in obesity for all generations of the English population and ``yo-yo'' dieting in the U.S.~data.
\end{abstract}

\textbf{Keywords:} cross-sectional data, longitudinal data, pooled data, Markov model, forecasting, BMI, marijuana

\section{Introduction}
\label{sec:intro}

The abundance of statistical surveys and censuses from past years invites new enhanced methods for studying various aspects of the composition and dynamics of populations. Gathered in different forms, as cross-sectional or longitudinal data, they provide information on large, independent or overlapping, sets of subjects drawn from a population and observed at several points in time. The first presents a snapshot of the population for quantitative and comparative analysis, while the latter tracks selected individuals, facilitating cohort and causal inferences. The cross-sectional data is often regarded inferior to the longitudinal one as it does not capture mechanisms underpinning observed effects. At the same time, however, it is oblivious to such problems as attrition, conditioning or response bias, while its much cheaper and faster collection procedure does not raise concerns about the confidentiality and data protection legislation. For these reasons, it is tempting to search for ways of employing it in the longitudinal analysis.

Making inferences about the population dynamics on the basis of severed longitudinal information gleaned from cross-sectional data requires suitable theoretical approach and modelling tools. Several methods proposed, e.g.~\cite{LeeJudgeZellner,Kalbfleisch1983,McCullagh,long97,dobson2008,goodman1959,king1997,PenubartiSchuessler,Moffitt1990,Moffitt1993,Collado1997,achenshively,Eisinga2008,Pelzer2002,Verbeek2000,Meng201496,BernsteinS16,hawkins}, are essentially based on regression techniques, concern cohort studies or special cases of modelling repeated cross-sectional data using Markov models. In this paper we present a generally applicable cross-sectional Markov (CSM) model for the transition analysis of survey data exploiting information from cross-sectional samples. While sharing the attractive features of classic Markov models, it avoids the practical problems associated with longitudinal data, and---due to its focus on population transfer rates between discrete states---it is particularly well adapted to microsimulation modelling. The presented framework provides a set of versatile and robust tools for the analysis and forecasting of data trends with applications in various disciplines, including epidemiology, economics, marketing and political sciences.

The following sections introduce the CSM model, presenting its detailed mathematical description and placing it in the context of statistics and information theory. We adapt it to analyse ageing cohorts and processes with memory, and demonstrate a regularisation method helping to avoid fallacious inferences. The framework is extended to describe incomplete longitudinal data, giving the possibility of fully exploiting all available information, e.g.~from combined surveys of different types. We also outline the model selection procedure required in its empirical applications as an integral part of any statistical data modelling. Finally, we test the developed methods and illustrate them by examples using actual demographic statistics, obtaining new interesting results.

\section{Theoretical description}
\label{sec:theor}

We begin by expounding the mathematical background and derivation of the CSM model. Sections\,\ref{sec:bayesian-analysis} and\,\ref{sec:mle} discuss it within the context of popular methods of statistical inference: Bayesian analysis and maximum likelihood estimation. In Sec.\,\ref{sec:kl-divergence} we give the information-theoretical interpretation of the maximum likelihood estimation used to determine the CSM model parameters. Sections\,\ref{sec:ageing_cohorts} and\,\ref{sec:memory}, respectively, present the time-inhomogeneous extension for simulations of ageing cohorts and the process memory. In the final section we describe model selection procedure to be used in the CSM framework applications.

\subsection{Model assumptions and parameters}
\label{sec:CSM}

We analyse a time series of observations of a certain characteristic (such as a risk factor, an exposure or a disease) in a studied population. The frequency and pattern of the characteristic in a surveyed representative sample is described by the distribution of a categorical variable $X$. The variable can take $N$ distinct values $k=0, 1, \dots, N - 1$ corresponding to different categories of the characteristic (e.g.~ranges of risk factor values or stages of a disease). The observations $\{ \obs{\vec{n}}_t \}$ are made at constant time intervals $t = 0, \dots, T - 1$. At each time point $t$, the sample of size $\obs{n}_t$ consists of groups of $\obs{n}_{kt}$ individuals assigned to category $k$, $\obs{\vec{n}}_t = (\obs{n}_{0t}, \dotsc, \obs{n}_{N-1,t})$; if data are missing for any $t$ then $\obs{\vec{n}}_t = 0$. Thus, the empirical distribution of $X_t$ is given by $\obs{p}_{kt} := \obs{n}_{kt} / \obs{n}_t$. Our goal is to find a model estimating population trends $p_{kt} := P(X_t = k)$ with their confidence intervals, and extrapolate the obtained results beyond the surveyed period.

When the survey follows the same individuals (a cohort) over time, as in longitudinal studies, a common approach is to use a (discrete-time) Markov model. It describes directly the stochastic dynamics of $X_{it}$, which is the value of variable $X$ for an observed individual $i$ at time $t$. The probability of $X_{it} = k$ conditioned on the value of $X_i$ at time $t-1$ is given by a constant $N \times N$ transition matrix $\mpi$ with elements $\pi_{kl}$ satisfying the constraints $\pi_{kl} \in [0,1]$ and $\forall_l \ \sum_{k = 0}^{N - 1} \pi_{kl} = 1$:
\begin{equation}
\label{eq:Markov}
P(X_{it} = k | X_{i,t-1} = l) = \pi_{kl} \ .
\end{equation}
However, if only the empirical distribution of $X_{t-1}$ in the investigated sample is known, then applying Bayes' theorem and $P(X_{i, t - 1} = l | \obs{\vec{p}}_{t-1} ) = \obs{p}_{l, t-1}$ we arrive at
\begin{equation}
\begin{split}
P(X_{it} = k | \obs{\vec{p}}_{t-1} ) &= \sum_l P(X_{it} = k\ \cap\ X_{i,t-1} = l | \obs{\vec{p}}_{t-1} ) =\\ =&\sum_l P(X_{it} = k | \obs{\vec{p}}_{t-1}\ \cap\ X_{i,t-1} = l ) \cdot P(X_{i,t-1} = l | \obs{\vec{p}}_{t-1} ) = \sum_l \pi_{kl} \obs{p}_{l,t-1} \ .
\end{split}
\label{eq:al-Markov}
\end{equation}
Taking the expectation values of both sides, we readily obtain
\begin{equation}
p_{kt} = \sum_{l = 0}^{N-1} \pi_{kl} p_{l,t-1} \ ,
\label{eq:cs-Markov}
\end{equation}
where $p_{kt} \in [0,1]$ and $\sum_{k = 0}^{N-1} p_{kt} = 1$, or $\vec{p}_t = \mpi \vec{p}_{t-1}$ in vector notation.

While Eq.\,\eqref{eq:Markov} involves tracking the same individual over time, Eq.\,\eqref{eq:cs-Markov} is conspicuously free from this assumption. It transforms the distribution of variable $X$ in a cross-sectional sample collected at time $t$ into the distribution of this variable in another such sample collected at time $t' > t$, using the transition matrix $\mpi$. By rearranging its terms, we can trace changes in the frequency of property $k$ between consecutive observations:
\[
\frac{n_{kt'}}{n_t'} - \frac{n_{kt}}{n_{t}} = \sum_{l = 0}^{N-1} ((\mpi^{t'-t})_{kl} - \delta_{kl}) \frac{n_{lt}}{n_{t}} \ ,
\]
where $n_t$ and $n_{kt}$ are sizes of the population and its categories and $\delta_{kl}$ is the Kronecker delta. Therefore, Eq.\,\eqref{eq:cs-Markov} can be the basis of a more robust dynamical model, which we propose in this article. Due to its mathematical construction, we call it the cross-sectional Markov model. It facilitates the description and projections of temporal trends of investigated characteristics in the population, its groups or ageing cohorts based on repeated cross-sectional data. The CSM model parameters: transition matrix $\mpi$ and initial distribution $\vec{p}_0$, are determined by fitting $\vec{p}_t = \mpi^t \vec{p}_0$ to the observed distributions $\obsvp_t$. In its extended version, the model utilises information collected in any form, including combined cross-sectional and (possibly incomplete) longitudinal data. It is achieved by maximising the log-likelihood of the data over $\mpi$ and $\vec{p}_0$. Detailed mathematical reasoning and procedures are described in the following sections.

\subsection{Bayesian analysis}
\label{sec:bayesian-analysis}

Estimation of standard Markov model parameters amounts to measuring the initial distribution of an investigated variable and counting frequencies  of transitions between its states at consecutive time steps. This straightforward procedure owes to the fact that continuous longitudinal trajectories provide full information about the dynamics of the observed process. In contrast, repeated cross-sectional data do not capture the individual transitions. To understand the implications of this difference for the CSM model, we perform a Bayesian analysis of the model parameters. We will demonstrate that the missing longitudinal information in repeated cross-sectional data results in a more complicated form of posterior distribution of $\vec{p}_0$ and $\mpi$. This creates additional challenges in its estimation and calculations of confidence intervals.

Within the Bayesian paradigm, the joint probability distribution of $\vec{p}_0$ and $\vec{\mpi}$ is inferred from observed data $I$. Before making any observations, we assume that all their values are equally probable, i.e.~their prior probability density function is $\rho( \vec{p}_0 , \mpi ) = 1$, which carries maximum entropy and therefore least extraneous information. Next, by conditioning it on the observation results we obtain the posterior distribution with the density $\rho( \vec{p}_0  , \mpi | I )$ representing best our state of knowledge about the distribution of model parameters given our prior knowledge and the data.

In a longitudinal study, we track changes of the investigated characteristic in a group of the same individuals over a period of time. From Bayes' theorem, we update the posterior with the likelihood of the value of variable $X_i$ for each individual $i$ added to the sample:
\begin{equation}
\rho( \vec{p}_0 , \mpi | I, X_{i,0} = k ) \sim p_{k,0} \, \rho( \vec{p}_0 , \mpi | I )
\label{eq:update-init}
\end{equation}
and at each time step $t$ of his or her longitudinal trajectory:
\begin{equation}
\rho( \vec{p}_0 , \mpi | I, X_{i,t} = k \land X_{i,t-1} = l ) \sim \pi_{kl} \, \rho( \vec{p}_0 , \mpi | I )\,,
\label{eq:update-trans}
\end{equation}
where $I$ is the previously observed data. From the above updating rules it follows that the final posterior joint density of the standard Markov model parameters is a product of Dirichlet distribution densities:
\begin{equation}
\label{eq:bayes-rho-cohort}
\rho( \vec{p}_0  , \mpi | I ) = \mathcal{D}_{\vec{\alpha}_0}(\vec{p}_0) \prod_{k=0}^{N-1} \mathcal{D}_{\vec{\alpha}_{k+1}}(\vec{\pi}_k) \ ,
\end{equation}
where $\mathcal{D}_{\vec{\alpha}}(\vec{x}) = {1}/{B(\vec{\alpha})}\,\prod_{k=0}^{N-1} x_k^{\alpha_k - 1}$ is an $N$-di\-men\-sio\-nal Dirichlet density with a parameter vector $\vec{\alpha}$ (i.e.~$\obs{n}_{0,k}=\alpha_{0,k}-1$ is the number of individuals in category $k$ at time 0 and $(\vec{\alpha}_{k+1})_l - 1$ is the overall number of transitions from category $k$ to $l$), $B(\vec{\alpha})$ $=$ ${\prod_{k=0}^{N-1} \Gamma(\alpha_k)}/{\Gamma(\sum_{k=0}^{N-1} \alpha_k)}$ and $\vec{\pi}_k$ is the $k$-th column vector of $\mpi$. Such Bayesian estimation of $\rho$ for longitudinal data can be easily implemented numerically.

In the case of cross-sectional data, the same non-informative prior is updated by observations of the form $X_t = k$, leading to the posterior density
\[
\rho( \vec{p}_0 , \mpi | I, X_t = k ) \sim p_{kt} \, \rho( \vec{p}_0 , \mpi | I ) = (\mpi^t \vec{p})_k \, \rho( \vec{p}_0 , \mpi | I ).
\]
For example, for $t = 1$ we obtain $\rho( \vec{p}_0 , \mpi | I, X_1 = k ) \sim \sum_{l = 0}^{N-1} \pi_{kl} p_{l, 0} \, \rho( \vec{p}_0 , \mpi | I)$. Thus, the posterior density of model parameters conditioned on the observed cross-sectional data is not a product of Dirichlet densities, like in the longitudinal case, but a mixture of such products:
\begin{equation}
\label{eq:bayes-rho-cs}
\rho( \vec{p}_0 , \mpi | \{ \obs{\vec{n}}_t \}) = \sum_{q=0}^{Q'-1} \beta_q \, \rho_q( \vec{p}_0 , \mpi | \{ \obs{\vec{n}}_t \}) 
= \sum_{q=0}^{Q'-1} \beta_q \mathcal{D}_{\vec{\alpha}_{q0}}(\vec{p}_0) \prod_{k=0}^{N-1} \mathcal{D}_{\vec{\alpha}_{q,k+1}}(\vec{\pi}_k) \,,
\end{equation}
where $\rho_q$ is the posterior distribution corresponding to each possible trajectory $\vec{k}_q$ realising the observed cross-sectional data, $Q'=\prod_{t=0}^T \sum_{k=0}^{N-1} 1_+(\obs{n}_{kt})$ with $1_+(a)$ equal 1 for $a>0$ and 0 otherwise, while $\sum_{q = 0}^{Q'-1} \beta_q = 1$ and $\beta_q \ge 0$. Consequently, the number of components of $\rho$ for cross-sectional data grows exponentially with the sample size and number of time steps, making its direct calculation impossible in practice.

The posterior distributions' densities~\eqref{eq:bayes-rho-cohort} and~\eqref{eq:bayes-rho-cs} are also characterised by different covariance structures. To demonstrate it, let $\vec{x}_j$ for $j = 0,\dotsc,N$ be defined as
\[
\vec{x}_j = \begin{cases}
\vec{p}_0 & j = 0 \\
\vec{\pi}_{j-1} & 1\le j \le N
\end{cases} \ .
\]
For longitudinal data, vectors $\vec{x}_j$ and $\vec{x}_{j'}$ are statistically independent for $j \neq j'$, due to the product structure of the posterior~\eqref{eq:bayes-rho-cohort}. However, for the cross-sectional posterior~\eqref{eq:bayes-rho-cs} we obtain $\mathbb{E}[ (\vec{x}_j)_k (\vec{x}_{j'})_{k'} ] = \sum_{q} \beta_q \mathbb{E}_q[ (\vec{x}_j)_k ] \mathbb{E}_q[ (\vec{x}_{j'})_{k'} ]$, where $\mathbb{E}_q$ denotes the expectation value calculated using $\rho_q( \vec{p}_0 , \mpi | \{ \obs{\vec{n}}_t \})$ and $\mathbb{E}[ (\vec{x}_j)_k ] = \sum_{q} \beta_q \mathbb{E}_q[ (\vec{x}_j)_k ]$. Hence, $\mathbb{E}[ (\vec{x}_j)_k (\vec{x}_{j'})_{k'} ] \neq \mathbb{E}[ (\vec{x}_j)_k ] \mathbb{E}[ (\vec{x}_{j'})_{k'} ]$ for $j \neq j'$, which indicates that $\vec{x}_j$ and $\vec{x}_{j'}$ are not statistically independent. This makes the analytic calculation of confidence intervals for estimated probabilities more difficult, unless one resorts to the ``delta method'' approach (postulating that the likelihood function is a multivariate Gaussian and expanding $\ln(p_{kt}/(1- p_{kt}))$ up to linear terms in $\vec{p}_0$ and $\mpi$), which can significantly underestimate their width. In general,  due to its complex structure the posterior density for cross-sectional data is broader than for longitudinal data, leading to wider confidence intervals.

For the above reasons (impracticable calculation of the posterior distribution and correlation of model parameters for cross-sectional data) we will use the maximum-likelihood method to obtain $\fit{\mpi}$ and $\fitvp_0$, point estimates of $\mpi$ and $\vec{p}_0$ (the next section), and bootstrapping for their confidence intervals (Appendix\,\ref{sec:bootstrapping}).

\subsection{Maximum likelihood estimation}
\label{sec:mle}

Given the difficulties in calculating the distribution of the CSM model parameters in the Bayesian framework, we attempt instead to find their ``best'' values $\fit{\mpi}$ and $\fitvp_0$ by maximising the log-likelihood $l[\vec{p}_0,\mpi] := \ln P( \{ \obs{\vec{n}}_t \} | \vec{p}_0,\mpi )$. For cross-sectional data we obtain
\begin{equation}
l_{\text{CS}}[\vec{p}_0,\mpi] = \ln \prod_{t=0}^{T-1} \prod_{i=0}^{\obs{n}_{t} - 1} P(X_t = k_{it} | \vec{p}_0,\mpi ) = \sum_{t=0}^{T-1} \obs{n}_{t} \sum_{k=0}^{N-1} \obsp_{kt} \ln p_{kt} = 
\sum_{t=0}^{T-1} \sum_{k=0}^{N-1} \obs{n}_{kt} ( \mpi^t \vec{p}_0 )_k \ .
\label{eq:ll-cs}
\end{equation}

The CSM framework can be extended to permit the analysis of incomplete longitudinal data (distorted by attrition or non-adherence). They can be represented as a set $\Theta = \{\vec t_i, \vec k_i\}$, consisting of $Q$ independent trajectories, that is, vectors $\vec k_i$ of $s$ consecutive categories measured in $\tau_i$ points in time $\vec t_i = \{ t_{i,s} \}$, where $i = 0, \dotsc, Q - 1$ and $s = 0, \dotsc, \tau_i - 1$. This notation enables us to describe trajectories starting and ending at different times and having gaps.
The likelihood of observing a trajectory $\vec{k}_{i} \in [0, N - 1]^{\tau_i}$ given $\vec{p}_0$ and $\mpi$ is
\begin{equation}
\label{eq:like-long}
P(\vec{k}_i| \vec{p}_0,\mpi) = (\mpi^{t_{i,0}} \vec{p}_0)_{k_{i,0}} \prod_{s=1}^{\tau_i - 1} \left(\mpi^{t_{i,s} - t_{i,s-1}}\right)_{k_{i,s}, k_{i,s-1}} \ .
\end{equation}
Thus, the log-likelihood of the whole dataset $\Theta$ is
\begin{equation}
\label{eq:ll-long}
l_{\text{L}}[\vec{p}_0,\mpi] = \sum_{i=0}^{Q-1} \left[ \ln [(\mpi^{t_{i,0}} \vec{p}_0)_{k_{i,0}}] + \sum_{s=1}^{\tau_i - 1} \ln [\left(\mpi^{t_{i,s} - t_{i,s-1}}\right)_{k_{i,s}, k_{i,s-1}}] \right] .
\end{equation}
It is easy to notice that for complete longitudinal trajectories, i.e.~$\vec t_{i} = (0, 1, \dotsc,$ $\tau_i - 1)$ for all $i$, the above result reduces to a simple product of Dirichlet distributions discussed in Sec.\,\ref{sec:bayesian-analysis}.

To model a mixture of cross-sectional and longitudinal data, we maximise the sum of log-likelihood functions~\eqref{eq:ll-cs} and~\eqref{eq:ll-long}, $l_{\text{CS}} + l_{\text{L}}$, over $\mpi$ and $\vec{p}_0$. In doing this, it is useful to know that although the cross-sectional part can be described by Eq.\,\eqref{eq:ll-long} for every $\tau_i=1$, it is more efficient to treat it as a separate term using Eq.\,\eqref{eq:ll-cs}. Numerically, the task requires solving a highly non-linear constrained optimisation problem.

It is worth mentioning that another kind of data, halfway between cross-sectional and longitudinal, arises from cohort studies in which we observe the same set of $\obs{n}$ individuals at each time $t$, but record only their number $\obs{n}_{kt}$ falling into given category $k$. Accordingly, to describe such anonymised longitudinal data, in Eq.\,\eqref{eq:ll-cs} we replace the unconditional probability predicted by the model $p_{kt} = (\mpi^t \vec{p}_0)_k$ with the one conditioned on previous observations $P(X_{it} = k|\obsvp_{t-1}) = (\mpi \obsvp_{t-1})_k$, as derived in Eq.\,\eqref{eq:al-Markov}, where we assume no gaps for simplicity. Historically considered intractable without approximations~\cite{Kalbfleisch1983}, the maximisation of the log-likelihood for anonymised longitudinal data is in fact easier than for cross-sectional one and can now be tackled with modern numerical algorithms and increased computational power.

In all the above cases, if we assume a ``flat'' prior like in Sec.\,\ref{sec:bayesian-analysis}, the maximum likelihood estimator of the model parameters is equivalent to their maximum posterior density estimator, bringing the maximum likelihood estimation procedure closer to the Bayesian analysis discussed in the previous section. This results from the fact that from Bayes' theorem, $\rho( \vec{p}_0,\mpi | \{ \obs{\vec{n}}_t \} ) = P( \{ \obs{\vec{n}}_t \} | \vec{p}_0,\mpi ) \rho( \vec{p}_0,\mpi ) / P(\{ \obs{\vec{n}}_t \})$, we obtain $\rho( \vec{p}_0,\mpi | \cdot ) \sim P( \cdot | \vec{p}_0,\mpi )$.

In specific situations the CSM model may not be able to uniquely recover the transition matrix because of the lack of information about individual transitions in repeated cross-sectional data, facing the threat of ecological inference fallacy~\cite{ecologicalfallacy}. For example, for $N = 2$ and a constant time series $\obsvp_t \equiv (1/2, 1/2)$, both $\fit{\pi}_{kl} = \delta_{kl}$ and $\fit{\pi}_{kl} = 1/2$ (maximum and minimum correlation between $X_t$ and $X_{t-1}$, respectively) are perfect solutions of Eq.\,\eqref{eq:cs-Markov}, and so is their convex combination. To steer the model estimation procedure towards a particular outcome, one can subtract from $l_{\text{CS}}$ a regularisation term $\Lambda_1 \sum_{k,l = 0}^{N-1} (\pi_{kl} - \Lambda_2 \delta_{kl})^2$, where $\Lambda_1 \ge 0$ is the regularisation strength and $\Lambda_2 \in [0, 1]$ specifies the type of required solution (from $\fit{\pi}_{kl} = \delta_{kl}$ for $\Lambda_2 = 1$ to $\fit{\pi}_{kl} = 1/2$ for $\Lambda_2 = 0$). Within the Bayesian framework, adding the penalty term of the above form amounts to replacing the uniform prior with $\rho( \vec{p}_0 , \mpi ) \sim \prod_{k,l=0}^{N-1} \varphi(\pi_{kl}; \Lambda_2 \delta_{kl},(2\Lambda_1)^{-1})$, where $\varphi(\cdot ; \mu, \varsigma^2)$ is the density of normal distribution $\mathcal N(\mu, \varsigma^2)$. Consequently, we solve maximum posterior rather than maximum likelihood, which are not the same in this situation. While realistic time-dependent data usually sufficiently constrains the solution space for $\fit{\mpi}$, some form of regularisation may be required at times, e.g.~for purely cross-sectional data exhibiting simple temporal trends.

\subsection{Connection to information theory}
\label{sec:kl-divergence}

To better understand the relation between the CSM model and other popular approaches to modelling repeated cross-sectional data (such as regression methods), it is helpful to couch it in terms of information theory. Within this framework we represent our procedure of finding the model parameters as a fitting of estimated probability distributions $\fitvp_t := \fit{\mpi}^t \fitvp_0$ to observed ones $\obsvp_t$. The norm of the estimation error which is being minimised by the calibration procedure is the Kullback--Leibler divergence~\cite{kullback1951,kullback1987},
\[
\DKL(\obsvp_t \Vert \vec{p}_t) := \sum_{k=0}^{N-1} \ln \left( \frac{\obsp_{kt}}{p_{kt}} \right) \obsp_{kt} = \sum_{k=0}^{N-1} \obsp_{kt} \ln \obsp_{kt} - \sum_{k=0}^{N-1} \obsp_{kt} \ln p_{kt}\,.
\]
This important information-theoretical measure corresponds to the amount of information lost when replacing the probabilities indicated directly by the data with our model estimates. Consequently, by minimising $\DKL(\obsvp_t \Vert \vec{p}_t)$ over $\mpi$ and $\vec{p}_0$ we ensure that our model exploits the maximum information from the data, without making any distributional assumptions about the error~\cite{kullback1959,Manning2008}. 

It is easy to show that maximising log-likelihood for cross-sectional data~\eqref{eq:ll-cs} over $\vec{p}_0$ and $\mpi$ is equivalent to minimising the weighted sum of Kullback--Leibler divergences $\DKL(\obsvp_t \Vert \vec{p}_t)$,
\begin{equation}
\label{eq:minDKL}
\min_{\vec{p}_0,\mpi} \sum_{t=0}^{T-1} \obs{n}_t \sum_{k=0}^{N-1} \ln
\left( \frac{\obsp_{kt}}{p_{kt}} \right) \obsp_{kt} \ .
\end{equation}

Similarly, in the simple case of complete longitudinal trajectories with equal lengths $\tau_i = T$, maximising their log-likelihood is equivalent to minimising the Kullback--Leibler divergence of two probability distributions over the space of all possible trajectories of length $T$. To demonstrate this, let $\obs{n}_{\vec{k}}$ be the number of observations of trajectory $\vec{k}$ in the dataset $\Theta$. Then, $\obsp_{\vec{k}} = \obs{n}_{\vec{k}} / Q$ is the distribution of different trajectories, which we want to approximate, an equivalent of $\obsp_{kt}$ in Eq.\,\eqref{eq:minDKL}. The Kullback--Leibler divergence to be minimised over $\vec{p}_0$ and $\mpi$ is
\[
\DKL(\obsvp_{\vec k} \Vert \vec{p}_{\vec k}) = -\sum_{\vec k} \obsp_{\vec{k}} \left( \ln P(\vec{k}| \vec{p}_0,\mpi) - \ln \obsp_{\vec{k}} \right) \,,
\]
where $P(\vec{k}| \vec{p}_0,\mpi)$ is given by Eq.\,\eqref{eq:like-long}. The first term in this formula is equal to minus log-likelihood of $\Theta$ given by Eq.\,\eqref{eq:ll-long}, which confirms our assertion.

\subsection{Modelling ageing cohorts based on cross-sectional data}
\label{sec:ageing_cohorts}

\begin{figure}[!htp]
\centering
\includegraphics[width=0.12\linewidth, angle=-90,trim=200px 200px 200px 200px]{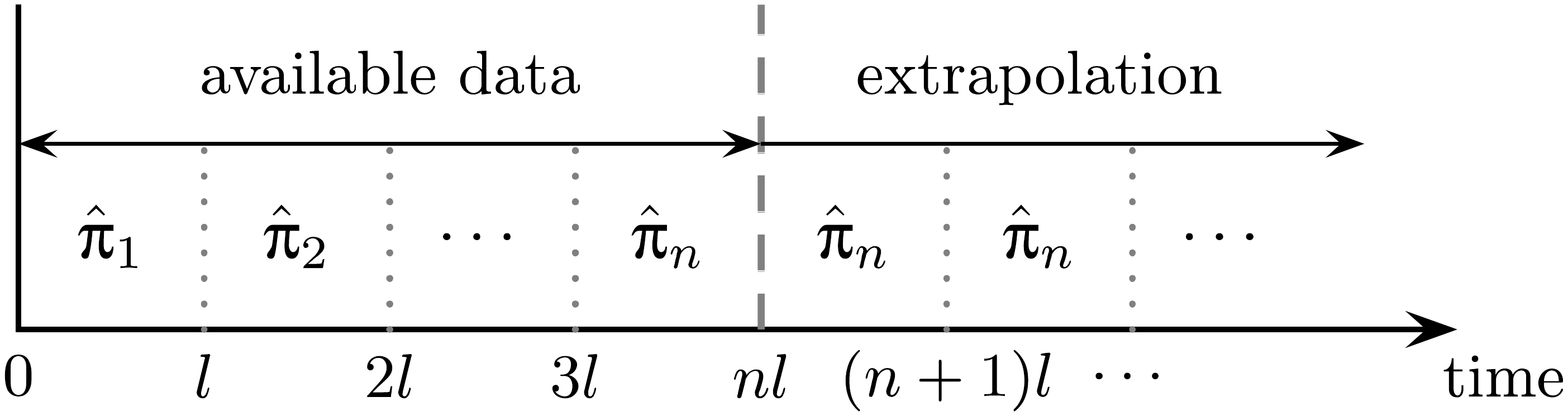}
				    \label{fig:cohortsim}
		    \caption{Schematic of the CSM model analysis for an ageing cohort.}
\end{figure}

The CSM model can be adjusted to analyse trends of characteristics in ageing cohorts based on repeated cross-sectional data or its mixture with longitudinal one. Figure\,\ref{fig:cohortsim} presents a schematic solution of this problem. It requires the estimation of CSM transition matrices $\fit{\mpi}$ and initial distributions $\fitvp_0$ for each age group and in all time periods covered by the available data. The age brackets are numbered from 1 to $n$ and assumed to have the same length $l$ for simplicity. Beginning with the initial state for the first bracket, we apply the transition matrix $\fit{\mpi}_1$ obtained for this bracket $l$ times, thus increasing the age of the cohort. This gives the first $l$ years of the CSM model fit $\fitvp$. Next, we switch to the second age bracket and apply the transition matrix $\fit{\mpi}_2$ to the current distribution $\fitvp_l$; we perform the operation $l$ times obtaining the fit for $l$ subsequent years. The procedure repeats until the last age bracket, at which point we keep on applying the last transition matrix $\fit{\mpi}_n$ until the end of the desired extrapolation period.

\subsection{Non-zero memory}
\label{sec:memory}

In discrete-time Markov models, the memory of the process is introduced by conditioning the next state of the variable $X$ on not just the current one, but also one or more preceding states. For example, to model a one-step memory we can replace Eq.\,\eqref{eq:Markov} with
\begin{equation}
P(X_{it} = k | X_{i,t-1} = l \wedge X_{i,t-2} = m) = \pi_{klm} \,,
\label{eq:Markov-2}
\end{equation}
with constraints $\forall_{l,m} \sum_{k=0}^{N-1} \pi_{klm} = 1$ and $\pi_{klm} \in [0, 1]$.

A similar procedure can be performed in the CSM framework by estimating the joint distribution of $X_t$ and $X_{t-1}$ based on repeated cross-sectional data. For this purpose, we define a random variable $Z_t := (X_t, X_{t-1})$, noting that $Z_t$ and $Z_{t-1}$ are always correlated since both depend on $X_{t-1}$. The distribution of $Z_t$ is denoted by $\vec{q}_{t}$; its dynamics are governed by a transition matrix $\mat\zeta$, the counterpart of $\mpi$. Hence, $\mat\zeta$ has the form $\zeta_{(k,l), (l',m)} = \delta_{ll'} \pi_{klm}$, where $\pi_{klm}$ satisfies the same constraints as in Eq.\,\eqref{eq:Markov-2}. It follows that $q_{(k,l),t} = \sum_{m=0}^{N-1} \pi_{klm} q_{(l,m), t-1}$. Since we do not observe the process $Z_t$ directly, we have to estimate the initial state $\vec{q}_0$ and transition matrix $\mat\zeta$ based on $\obsvp_t$. It requires reducing the dimension of the distributions $\vec{q}_t = \mat\zeta^t\vec{q}_0$ from $N^2$ to $N$ to the end that $\vec{p}_t = R[\vec{q}_t]$, where the reduction operator is defined as
\[
R[\vec{q}_t]_k := \sum_{l=0}^{N-1} q_{(k,l),t} \ .
\]
The estimates $\fit{\mat\zeta}$ and $\fitvq_0$ are found by minimising the total weighted Kullback--Leibler distance,
\[
\min_{\mat\zeta, \vec{q}_0} \sum_{t=0}^{T-1} n_t \DKL(\obsvp_t \Vert R[\mat{\zeta}^t \vec{q}_0])\ .
\]
The above procedure easily extends to models with longer memory and has a straightforward numerical implementation.

In the general case of a mixture of cross-sectional and longitudinal data, introducing the memory length $\lambda$ leads to a more complicated form of the likelihood function~\eqref{eq:like-long}. Let $G = [0,N-1]^{\lambda+1}$ be the set of values of $Z_t = (X_t, \dotsc, X_{t-\lambda})$, containing all possible continuous sequences of states of variable $X_t$ spanning the memory length $\lambda$. The transition matrix $\mpi$ is indexed by such a sequence $\vec\xi \in G$ (representing a trajectory over the time interval $[t-\lambda,t]$) and a state $k \in [0,N-1]$ at $t+1$, that is $\pi_{k\vec\xi}$. We define $\sigma(\vec{k}_i)_t \subset G$ as the set of only those sequences for which the event $Z_t = \vec\xi$ is not contradicted by the observed trajectory $\vec{k}_i$. For example, we model a variable with $N = 3$ states and memory $\lambda = 2$ based on three consecutive longitudinal measurements, one of which is missing: $(X_0 = 1, X_1 = \text{unknown}, X_2 = 2)$; then $\sigma(k_i)_2 = \{(2, 0, 1), (2, 1, 1), (2, 2, 1)\}$. The final form of the likelihood function for $\vec{k}_i$ starting at $t_{i,0}$ and ending at $t_{i,\tau_i-1}$ is
\[
P(\vec{k}_i| \vec{q}_0, \mat{\xi}) = \sum_{\xi{\tau_i-1} \in \sigma(\vec{k}_i)_{t_{\tau_i-1}}} \dotsc \sum_{\xi_0 \in \sigma(\vec{k}_i)_{t_{q,0}}} \zeta_{\xi_{\tau_i-1},\xi_{\tau_i-2}} \dotsc \zeta_{\xi_{1},\xi_{0}} q_{\xi_0,0}  \ ,
\]
where $\zeta_{\vec\xi',\vec\xi} = \delta_{\xi'_1,\xi_0} \dotsm \delta_{\xi'_\lambda,\xi_{\lambda-1}} \pi_{\xi'_0\vec\xi}$. A fully-specified trajectory (without gaps) will always have only a single element in every set $\sigma(\vec{k}_i)_t$ for $t_{q,0} + \lambda \le t \le t_{q,\tau_i-1}$.

The error estimation by bootstrapping is performed identically as in the case of zero memory.

\subsection{Selecting the best model}
\label{sec:model-selection}

An integral part of all statistical work with data is choosing a suitable model for their analysis. This section invokes the most popular model selection techniques and incorporates them in the proposed framework. They will enable us to quantitatively compare the performance of the CSM model variants and other methods. On this basis, we will decide which model best explains the mechanisms underlying the data, providing the most accurate and stable forecasts.

Assuming that the calibration procedure converges numerically, one can always improve the fit by increasing the model complexity (e.g.~extending the CSM memory length). However, indiscriminately adding new model parameters leads to overfitting. To strike a balance between these two factors, we calculate Akaike Information Criterion (AIC)~\cite{Akaike}
\begin{equation}
\text{AIC} = 2 k - 2 l_\text{max}
\label{eq:aic}
\end{equation}
and Bayesian Information Criterion (BIC)~\cite{Schwarz1978}
\begin{equation}
\text{BIC} = k (\ln n - \ln 2 \pi) - 2 l_\text{max}\,,
\label{eq:bic}
\end{equation}
where $k$ is the number of model degrees of freedom and $n$ is the sample size (i.e.~the number of collected surveys for cross-sectional and observed trajectories for longitudinal data).\footnote{Using the number of subjects as the longitudinal sample size is a very conservative assumption, which can underestimate the effective size. However, the CSM framework is amenable to this basic approach, as we will demonstrate in Sec.\,\ref{sec:application}. Additionally, since typical applications of the CSM model concern large $n$, we neglect the commonly used small-sample correction to AIC.} Both criteria are constructed as a penalty for the complexity of a candidate model minus twice the maximised log-likelihood $l_\text{max}$ value, but they are derived from different mathematical perspectives. AIC intends to minimise the Kullback--Leibler distance between the true data-generating probability density and that predicted by the candidate model, while BIC seeks for the model with maximum posterior probability given the data. Consequently, they do not always select the same ``best'' candidate. In particular, BIC assures consistency (for very large sample size it will choose the correct model with probability approaching 1), while AIC aims at optimality (as more benevolent to complexity, it leads to a lower variance of the estimated model parameters, especially if their true values are close to those from oversimplified models)~\cite{yang2005,ModelSelection}. The difference between the criteria becomes evident with growing sample size: AIC allows additional model parameters to describe the new data (increasing the predictive accuracy if it represents new information, or overfitting if it totes mostly noise and outliers), while BIC is more stringent and favours smaller models (with fewer parameters). 

While the information criteria focus on accuracy and parsimony, cross-validation analysis is a natural and practical way of assessing the predictive performance and robustness of the model~\cite{arlot2010}. As a realisation of out-of-sample testing, it does not rely on analytical approximations but exact algorithms and provides a stronger check against overfitting. We will use its most common variants: leave-one-out (LOOCV) and $k$-fold cross-validation ($k$FCV), which under certain conditions behave similarly to AIC and BIC, respectively~\cite{yang2005}, as well as time series cross-va\-li\-da\-tion (TSCV).

The application of the above techniques is straightforward when working with cross-sectional data. In the case of LOOCV, given a time sequence of $T$ observations $\{ \obsvp_t \}$, for each $t \in [1, T)$ we calibrate the CSM model to all data points but $\obsvp_t$ and next use the obtained model to calculate the approximate value of the omitted point, ${\fitvp_t\tilde{}}^{\,t}$. The Kullback--Leibler divergence $D^\text{LOOCV}_t := \DKL(\obsvp_t \Vert {\fitvp_t\tilde{}}^{\,t})$ measures the error with which the model recovers $\obsvp_t$. The sum of $D^\text{LOOCV}_t$ over $t$, $D^\text{LOOCV}$, defines a measure of the model error which additionally discourages overfitting.

The $k$FCV consists in randomly partitioning, one or several times, the sequence of observations into $k$ subsets of equal size and performing the leave-one-out analysis on the subsets, adding the errors up for all such folds and averaging the sums over the partitionings. Since in each fold we leave out more data points, $k$FCV is a stronger test for overfitting than LOOCV.

Both LOOCV and $k$FCV are sufficient for testing the model's approximation and forecasting abilities, however they may fail at the latter in the setting of highly autocorrelated data~\cite{Hart1990}. To assess the CSM model performance in extrapolating this type of time series we need a specialised method such as TSCV~\cite{Hart1994}. In this approach, we fit the model to first $T'$ time points $t = 0,\dotsc,T'-1$, where $T' \ge 2$, and use it to predict the probability distribution for $T'$, $\fitvp_{T'}^{\text{TS}}$. The total TSCV error is the sum of Kullback--Leibler divergences $\DKL(\obsvp_{T'} \Vert \fitvp_{T'}^{\text{TS}})$ over $T' \in [2, T)$.

When working with data containing longitudinal information, we consider each individual trajectory (complete or with gaps) to be an independent observation. Hence, we perform $k$FCV by dividing the set of observed trajectories $\Theta$ (see Sec.\,\ref{sec:mle}) into $k$ subsets. Due to the multitude of trajectories, we are able to perform only a limited number of iterations of the procedure, namely 30. We perform TSCV by truncating the trajectories to test the stability of extrapolation results. The error of extrapolation from $T'-1$ to $T'$ is measured as a difference of log-likelihoods of trajectories truncated at $T'-1$ and $T'$, calculated using the model fitted to trajectories truncated at $T'-1$.

In the presence of a regularisation term, AIC and BIC are slightly less presumptive as instead of maximum log-likelihood, they use the logarithm of maximum posterior distribution, which is no longer equivalent, as shown at the end of Sec.\,\ref{sec:mle}. Since applying a penalty amounts to imposing constraints on the model parameters, one can improve the information criteria estimations by changing the number of degrees of freedom used to calculate them~\cite{Shedden2008}. In our case, however, the sample size (to which the log-likelihood function is proportional) and the memory length parameters dominate the impact of the term on maximised log-likelihood, and hence we can omit this correction. The cross-validation metrics are free from this problem as they use the Kullback-Leibler divergence to score the models.

Lastly, a relevant measure of the model's robustness and suitability for analysing a particular dataset is whether the extrapolated trends it produces behave reasonably and stably.

\section{Tests and empirical illustration}
\label{sec:application}

In this section we test the proposed framework using synthetic data (Sec.\,\ref{sec:synth}) and demonstrate its practical applications to a selection of repeated cross-sectional and incomplete longitudinal samples collected from real-life observational studies (Secs.\,\ref{sec:bmi}--\ref{sec:longit-bmi}). We employ different variants of the CSM model (with memory and regularisation) and compare them with the popular multinomial logistic regression (MLR)~\cite{McCullagh,long97,dobson2008} using the maximum likelihood estimator in both cases. The model selection procedure enables us to choose the best model, representing the most trustworthy and reliable statistical description and forecasting tool for the investigated problem.

All models were implemented in C++11, employing Open Source optimisation library NLopt~\cite{nlopt}
, automatic differentiation library Sacado~\cite{sacado} and linear algebra library Eigen~\cite{eigen}.

\subsection{Synthetic data}
\label{sec:synth}

To test the CSM framework, we use synthetic data consisting in a set of 1000 longitudinal trajectories, each of length 30, generated by a 3-dimensional Markov process with a one-period memory, as in Eq.\,\eqref{eq:Markov-2}. The initial state is
\begin{equation}
P_\text{input}(X_0 = k \wedge X_{-1} = l) := \begin{pmatrix}
0.08 & 0.14 & 0.08 \\
0.14 & 0.08 & 0.08 \\
0.10 & 0.10 & 0.20
\end{pmatrix}_{kl}
\label{eq:synth-initst-gen}
\end{equation}
and the transition matrix is described by Table\,\ref{tab:synth-pi}.
\begin{table}[h]
\centering
{\small
\begin{tabular}{|l|l|l|l|l|}
\hline
\multirow{2}{*}{$m$} & \multirow{2}{*}{$l$} & \multicolumn{3}{|c|}{$(\mpi_\text{input})_{klm} = P(X_{t+1} = k | X_{t} = l \wedge X_{t-1} = m)$} \\
\cline{3-5}
& & $k=0$\qquad\quad\ \,\ & $k=1$\qquad\quad\ \,\ & $k=2$\\
\hline
0 & 0 & 0.8 & 0.1 & 0.1\\
0 & 1 & 0.15 & 0.75 & 0.1\\
0 & 2 & 0.18 & 0.7 & 0.12\\
1 & 0 & 0.8 & 0.19 & 0.01\\
1 & 1 & 0.03 & 0.94 & 0.03\\
1 & 2 & 0.01 & 0.14 & 0.85\\
2 & 0 & 0.1 & 0.6 & 0.3\\
2 & 1 & 0.2 & 0.5 & 0.3\\
2 & 2 & 0.09 & 0.9 & 0.01\\
\hline
\end{tabular}}
\caption{Transition matrix coefficients used to generate synthetic data.}
\label{tab:synth-pi}
\end{table}

We use variants of the CSM model with different memory length $\lambda$ to analyse the longitudinal data and its reduction to cross-sectional form. The latter is obtained by calculating the count of category $l$ at time $t$ given a set of trajectories $\Theta$ (see Sec.\,\ref{sec:mle}),
\begin{equation}
\obs{n}_{lt} = \sum_{i=0}^{Q - 1} \sum_{q=0}^{\tau_i - 1} 1_{\{t\}}(t_{i,q}) 1_{\{l\}}(k_{i,q}) \ .
\label{eq:reduction}
\end{equation}
The calibration procedure, detailed in Secs.\,\ref{sec:mle} and\,\ref{sec:memory}, employs the log-likelihood~\eqref{eq:ll-long} for the longitudinal case CSM($\lambda$)$_\text{L}$ and \eqref{eq:ll-cs} for the reduced, cross-sectional case CSM($\lambda$)$_\text{CS}$. These two cases will give us an insight into the model's behaviour depending on the nature of the analysed problem.

\begin{figure}[h]
\centering
\includegraphics[width=0.67\linewidth, trim=0px 10px 0px 0px]{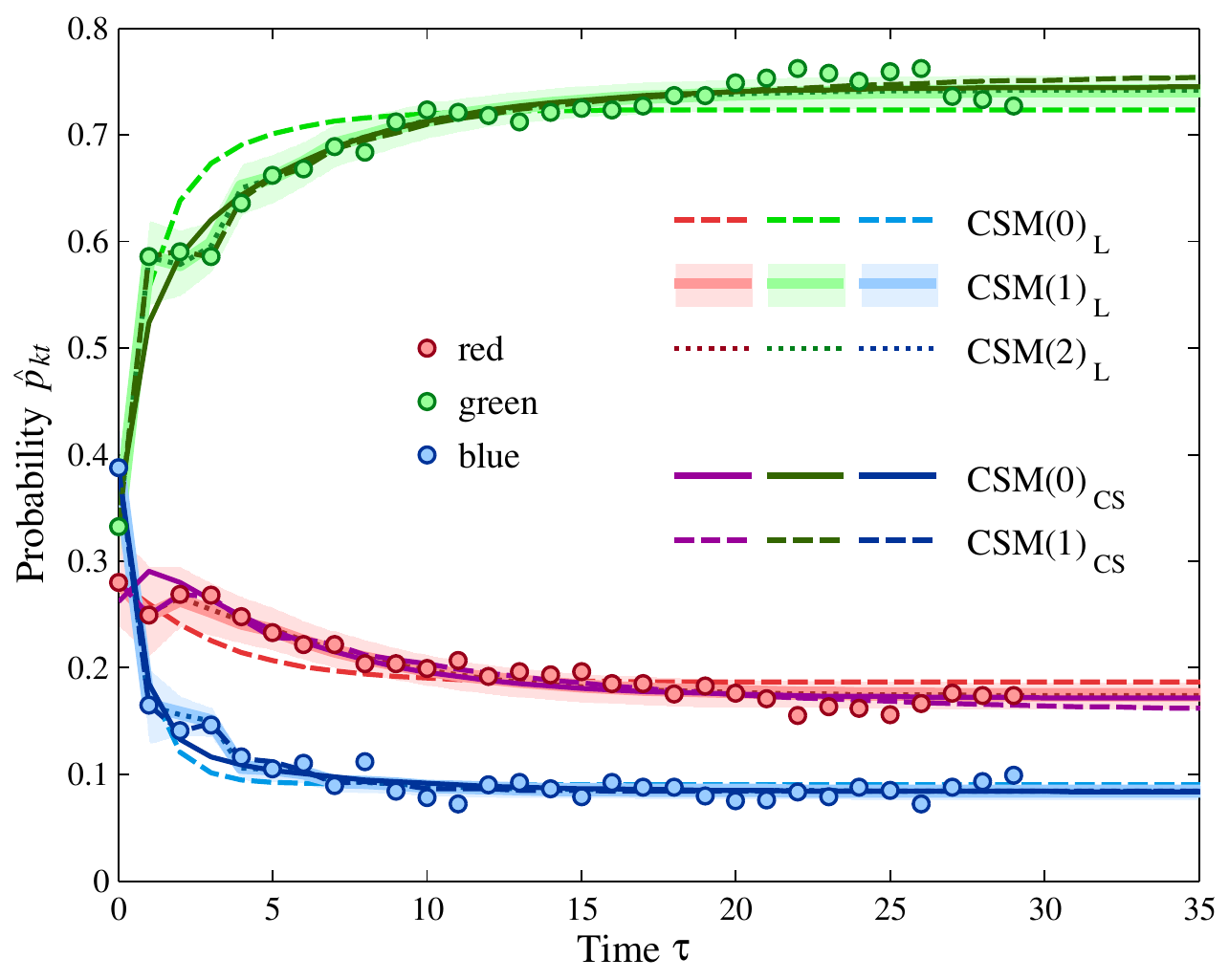}
\caption{The CSM model results for the synthetic data generated by the 3-state Markov process. The CSM($\lambda$)$_\text{L}$ and CSM($\lambda$)$_\text{CS}$ models have been calibrated to longitudinal and (reduced) cross-sectional data, respectively, with different memory lengths $\lambda$. The 95\% confidence intervals are shown for the best model, CSM$(1)_\text{L}$ (see Table\,\ref{tab:synth-longit}).}
\label{fig:synth}
\end{figure}

Figure \ref{fig:synth} presents the obtained fit and extrapolation results. All models, except CSM(0)$_\text{L}$, recover very well the observed distribution trends. Their long-term extrapolations converge towards the common steady state, defined by the equation $\mpi \vec{p} = \vec{p}$, signifying the stability of the CSM framework.

The bootstrapped confidence intervals, displayed for CSM(1)$_\text{L}$ in Fig.\,\ref{fig:synth}, have comparable widths for all models. To test the validity of the bootstrap procedure (Appendix\,\ref{sec:bootstrapping}), we compared them with confidence intervals obtained from a simulation, in which we generated new samples using $P_\text{input}$ and $\pi_\text{input}$ for each model calibration, instead of resampling the original one. The resulting confidence intervals are very similar to the bootstrapped ones, with the mean absolute lower/upper bound difference of just 0.0014.

We perform the model selection procedure, outlined in Sec.\,\ref{sec:model-selection}, to verify our results and pick the best candidate for the problem description. Table\,\ref{tab:synth-longit} (top) summarises the longitudinal case. The $\Delta$AIC and $\Delta$BIC scores, as well as $\Delta$LOOCV, $\Delta$\textit{k}FCV and $\Delta$TSCV residuals are reported relative to the best model according to the respective statistic (i.e.~the model with a zero in the corresponding column). By rule of thumb, a difference of more than 10 in a BIC value is considered to be strongly relevant~\cite{KassRaftery}, which we can also apply to other metrics as they depend on the maximum log-likelihood in a similar way. The variant of the model with memory length of one period, CSM(1)$_\text{L}$, is unequivocally the best. The remaining candidates, CSM(0)$_\text{L}$ and CSM(2)$_\text{L}$, contain too few or too many parameters to fit the available amount of data, leading to under- and overfitting, respectively. We conclude that the CSM framework correctly recovers the properties of the data-generating process and is congruent with standard model selection techniques.

\begin{table}[h]
\centering
\small{
\begin{tabular}{|l|c|c|c|c|c|c|}
\hline
Model & DOF & K-L div. & $\Delta$AIC & $\Delta$BIC & $\Delta$\textit{k}FCV & $\Delta$TSCV \\
\hline
CSM(0)$_\text{L}$ & 8 & 16290.7 & 5106.4 & 5052.5 & 2551.8 & 2311.3 \\
CSM(1)$_\text{L}$ & 26 & 13718.9 & 0 & 0 & 0 & 0 \\
CSM(2)$_\text{L}$ & 80 & 13706.7 & 96.2 & 249.4 & 159.3 & 512.5 \\
\hline\hline
Model & K-L div. & $\Delta$AIC & $\Delta$BIC & $\Delta$LOOCV & $\Delta$\textit{k}FCV & $\Delta$TSCV \\
\hline
CSM(0)$_\text{CS}$ & 32.28 & 0 & 0 & 0 & 0 & 0 \\
CSM(1)$_\text{CS}$ & 14.77 & 1.01 & 117.45 & 166.73 & 546.49 & 38.15 \\
\hline
\end{tabular}}
\caption{Model selection results for synthetic data: original longitudinal version (top panel) and their reduced cross-sectional form (bottom panel). For \textit{k}FCV, $k=5$ and the averaging is performed over 300 iterations.}
\label{tab:synth-longit}
\end{table}

Table\,\ref{tab:synth-longit} (bottom) presents the results for the cross-sectional case. Reducing longitudinal trajectories to cross-sectional distributions removed most information about the memory of the process. Consequently, we need less parameters to describe the data and the best model is the memory-less CSM(0)$_\text{CS}$.

Lastly, we compare the estimates of the initial state
\[
\fit{P}_{\text{CSM}_\text{L}(1)}(X_0 = k \wedge X_{-1} = l) := \begin{pmatrix}
0.08 & 0.15 & 0.04 \\
0.19 & 0.08 & 0.07 \\
0.04 & 0.09 & 0.25
\end{pmatrix}_{kl}
\]
and transition matrix (Table\,\ref{tab:synth-pi-calibr})
\begin{table}[h]
\centering
\small{
\begin{tabular}{|l|l|l|l|l|}
\hline
\multirow{2}{*}{$m$} & \multirow{2}{*}{$l$} & \multicolumn{3}{|c|}{$(\fit{\mpi}_{\text{CSM}_\text{L}(1)})_{klm}$} \\
\cline{3-5}
& & $k=0$ & $k=1$ & $k=2$\\
\hline
0 & 0 & 0.78 & 0.12 & 0.1\\
0 & 1 & 0.15 & 0.74 & 0.11\\
0 & 2 & 0.18 & 0.73 & 0.09\\
1 & 0 & 0.767 & 0.215 & 0.018\\
1 & 1 & 0.03 & 0.94 & 0.03\\
1 & 2 & 0 & 0.09 & 0.91\\
2 & 0 & 0.11 & 0.6 & 0.29\\
2 & 1 & 0.2 & 0.5 & 0.3\\
2 & 2 & 0.09 & 0.9 & 0.01\\
\hline
\end{tabular}}
\caption{Transition matrix coefficients obtained for longitudinal synthetic data from the best model CSM$_\text{L}$(1), $(\fit\mpi_{\text{CSM}_\text{L}(1)})_{klm} = P(X_{t+1} = k | X_{t} = l \wedge X_{t-1} = m)$.}
\label{tab:synth-pi-calibr}
\end{table}
obtained from the best model CSM$_\text{L}$(1) to $P_\text{input}$ and $\mpi_\text{input}$, demonstrating that the calibration procedure accurately recovers the true parameters of the data-generating process. In particular, the Frobenius distance of the transition matrices is $\lVert \fit\mpi_{\text{CSM(1)}_\text{L}} - \mpi_\text{input} \rVert = 0.00039$. By comparison, for CSM(1)$_\text{CS}$ calibrated to the reduced data, devoid of longitudinal information, the corresponding value is predictably higher and equals $\lVert \fit\mpi_{\text{CSM(1)}_\text{CS}} - \mpi_\text{input} \rVert = 1.75$. This affords an additional confirmation of the correct behaviour of employed numerical procedures.

\subsection{Cross-sectional BMI data on the English population}
\label{sec:bmi}

We employ the CSM framework to analyse the repeated cross-sectional data on the Body Mass Index (BMI [kg/m$^2$]) collected by the Health Survey for England (HSE) in years 1993--2013~\cite{bmi}. Twenty independent samples consist in between 3851 and 15303 persons aged 18 and older, each assigned to one of three BMI categories: normal weight (NW) with BMI~$\leq 25$, overweight (OW) with $25 <$~BMI~$\leq 30$ and obese (OB) with BMI~$> 30$. We investigate the BMI trends in the whole English population and in selected birth cohorts.

The tested variants of the CSM model include the memory-less CSM(0), the 1-year memory CSM(1), as well as CSM(0)$_\text{reg}$ penalising jumps by two BMI categories in one year (since a person's BMI changes continuously through the ordered categories, it is reasonable to expect that such a jump is less likely than remaining in the same or moving to the neighbouring category). The regularisation helps us to obtain realistic conditional probabilities of transitions by constraining the solution space for the $\mpi$ matrix, helping to avoid the ecological fallacy, as explained in the final paragraph of Sec.\,\ref{sec:mle}. Accordingly, we subtract from the log-likelihood $l[\hat{\vec{p}}_0, \hat{\pi}]$ the penalty term
\[
\Lambda \sum_{k,l=0}^{N-1} d_{kl} \pi_{kl}^2 \ ,\ \text{where}\ 
d_{kl} = \begin{cases}
1 &\text{for}\ \ |k - l| > 1 \\
0 &\text{for}\ \ |k - l| \le 1\ .
\end{cases}
\]
The above procedure is an example of Tikhonov regularisation~\cite{Hoerl} and leads to a new model, which needs to be compared with other candidates using the information criteria and cross-validation, as described in Sec.\,\ref{sec:model-selection}. We test several orders of magnitude of the regularisation strength $\Lambda \in [10, 10^5]$.

\begin{figure}[!htp]
\centering\includegraphics[width=0.67\linewidth, trim=0px 10px 0px 0px]{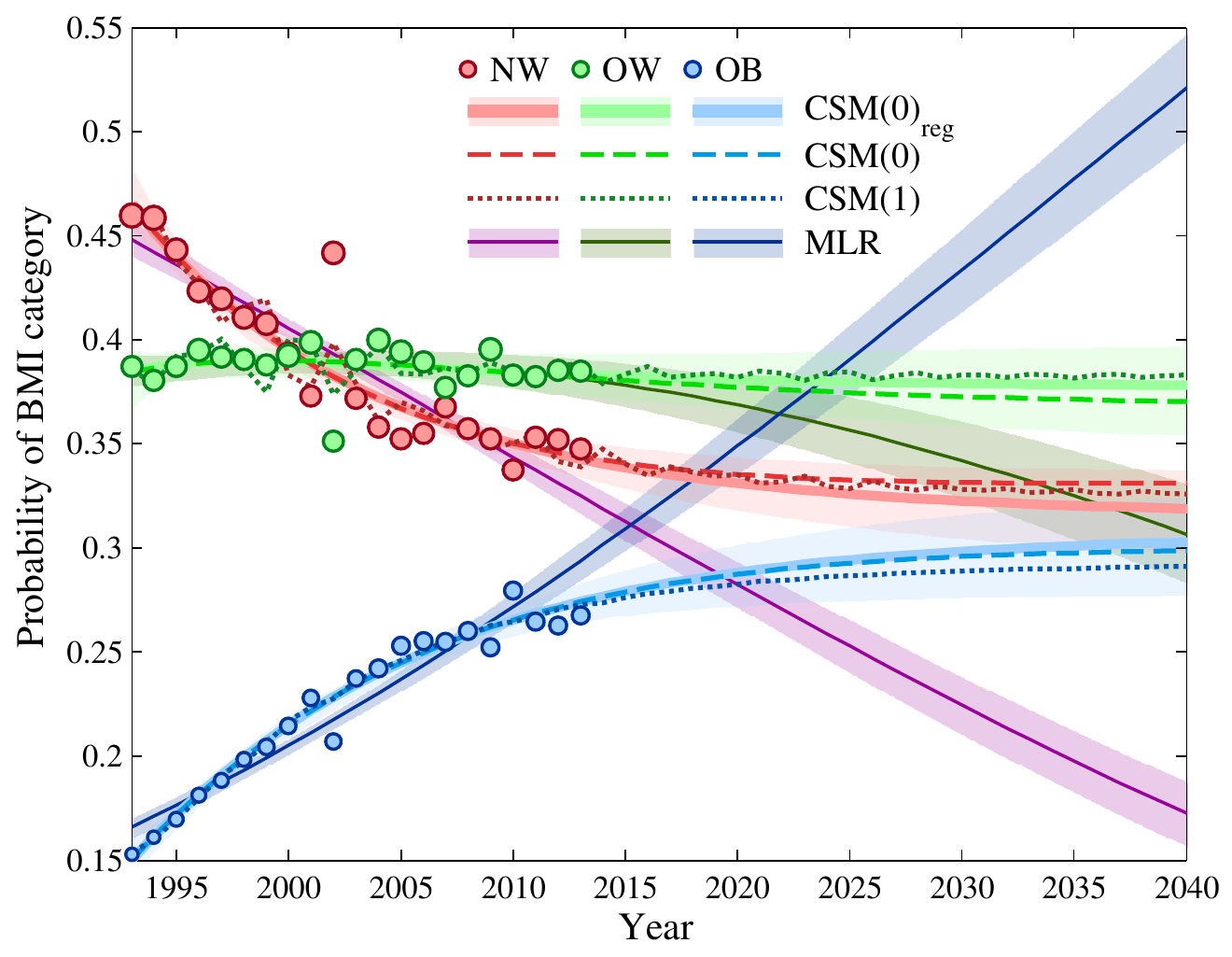}
\caption{Fits and projections of BMI trends (with 95\% confidence intervals) in adult English population obtained by CSM($\lambda$) models (variants with regularisation $\Lambda = 10^3$ and without, and memory length of $\lambda$ years) and MLR, based on repeated cross-sectional data from HSE for years 1993--2013 (marker area is proportional to the category count).}
\label{fig:population}
\end{figure}

The projected BMI trends in the adult English population are presented in Fig.\,\ref{fig:population}. All CSM models anticipate an increasing prevalence of excessive body weight: the fraction of obese persons increases, while that of normal-weight ones decreases; the overweight part exhibits a very mild decrease. The long-term trends flatten converging towards a steady state consisting in 32\% NW, 38\% OW and 31\% OB for CSM(0)$_\text{reg}$ with $\Lambda = 10^3$. Other CSM models attain a similar infinite time limit (despite visible overfitting in the CSM(1) case), proving the stability of the CSM framework. The regularisation decreases the variance of model parameters and thus narrows the confidence intervals as compared to the penalty-free versions (results for other regularisation strengths do not differ significantly). By comparison, the MLR method anticipates a much faster (supralinear) persistent growth of the obese group with a simultaneous shrinkage of normal weight and overweight groups, converging to an entirely obese population, which is rather imposed by the mathematical structure of the method than founded on the data.

The regularisation transforms the transition matrix implied by the CSM(0) model into a substantially different form,
\begin{equation}
\fit\mpi_{\text{CSM}(0)} = \begin{pmatrix}
0.957 & 0 & 0.048 \\
0.043 & 0.945 & 0.021 \\
0 & 0.055 & 0.932
\end{pmatrix} \text{and }\fit\mpi_{\text{CSM}(0)_\text{reg}} = \begin{pmatrix}
0.911 & 0.072 & 0.004 \\
0.089 & 0.873 & 0.065 \\
0 & 0.055 & 0.931
\end{pmatrix} \text{for $\Lambda = 10^3$},
\label{eq:UKBMIpi}
\end{equation}
reflecting the previously invoked problem of multiple solutions for purely cross-sectional data and the subsequent need for regularisation. Specifically, whereas both models agree that BMI of the majority of population does not change, they achieve the observed trends (Fig.\,\ref{fig:population}) in different ways. In CSM(0), normal weight and overweight persons (two first columns of $\fit\mpi_{\text{CSM}(0)}$) are only allowed to put on weight. To balance out this effect, the obese ones (third column) are more likely to reduce their BMI by two categories in one year. The regularised CSM(0)$_\text{reg}$ avoids this unrealistic effect and attributes most BMI changes to transitions between neighbouring categories. For instance, based on its predictions, the England's 45 million adult population currently consists in 34\% NW, 38\% OW and 28\% OB persons. By the following year the obesity rate will increase by 0.2\% (88 thousand people), the number by which the normal weight group will decline. The surprisingly rich dynamics behind this process involves 9\% of normal weight persons becoming overweight, 5.6\% of overweight persons turning obese and 7\% reducing their weight to normal, as well as 6.5\% of obese persons dropping to the overweight category. Increasing the regularisation parameter $\Lambda$ does not affect the calibrated transition matrices significantly.

Model selection procedure summarised in Table\,\ref{tab:BMIerrors} provides rigorous evaluation of the considered candidates. Comparison of the approximation errors shows that CSM models fit the data better than MLR, CSM(1) being the best. However, we need to appreciate the fact that it has more free parameters than the others. Increasing the number of degrees of freedom (DOF) almost always results in a better fit, but can also make the model less stable and more sensitive to noise and outliers. This is one of the reasons why the minimised error value or visual inspection of the fit are insufficient to validate the model, and we always need to resort to comprehensive statistical methodology. Accordingly, we find that AIC is dominated by the maximum log-likelihood term due to the large sample size, and hence picks the most complex model, CSM(1), as the best. At the same time, BIC prefers simpler models, CSM(0) and CSM(0)$_\text{reg}$, due to its stronger penalty for the number of parameters. Since we expect to see simple trends in the data, we interpret the AIC result as overfitting, which can indeed be observed in Fig.\,\ref{fig:population}. Furthermore, the cross-validation procedure unequivocally selects both memory-less CSM models. We conclude that the best candidates for studying temporal trends of BMI distributions are CSM(0) and CSM(0)$_\text{reg}$, the latter additionally yielding a realistic description of individual transitions between BMI categories.

\begin{table}[!htp]
\centering
{\small
\begin{tabular}{|l|c|c|c|c|c|c|c|}
\hline
Model  & DOF & K-L div. & $\Delta$AIC & $\Delta$BIC & $\Delta$LOOCV & $\Delta$\textit{k}FCV & $\Delta$TSCV \\
\hline
CSM(0)$_\text{reg}$ ($\Lambda = 10^3$) & 8 & 87 & 21.5 & 0.7 & 0 & 0 & 0 \\
CSM(0) & 8 & 90 & 20.9 & 0 & 1.9 & 1.4 & 2 \\
CSM(1) & 26 & 62 & 0 & 130 & 814 & 14638	& 42 \\
MLR & 4 & 160 & 152 & 98	& 80	& 83 & 59 \\
\hline
\end{tabular}}
\caption{Model selection results for the HSE dataset. For \textit{k}FCV, $k=5$ and the averaging is performed over 300 iterations. The results for CSM(0)$_{\text{reg}}$ with other values of $\Lambda \in [10, 10^5]$ differ by at most 3.3 (TSCV) and 0.64 (BIC, AIC).}
\label{tab:BMIerrors}
\end{table}

\begin{figure}[!htp]
\centering
\includegraphics[width=0.67\linewidth, trim=0px 0px 0px 0px]{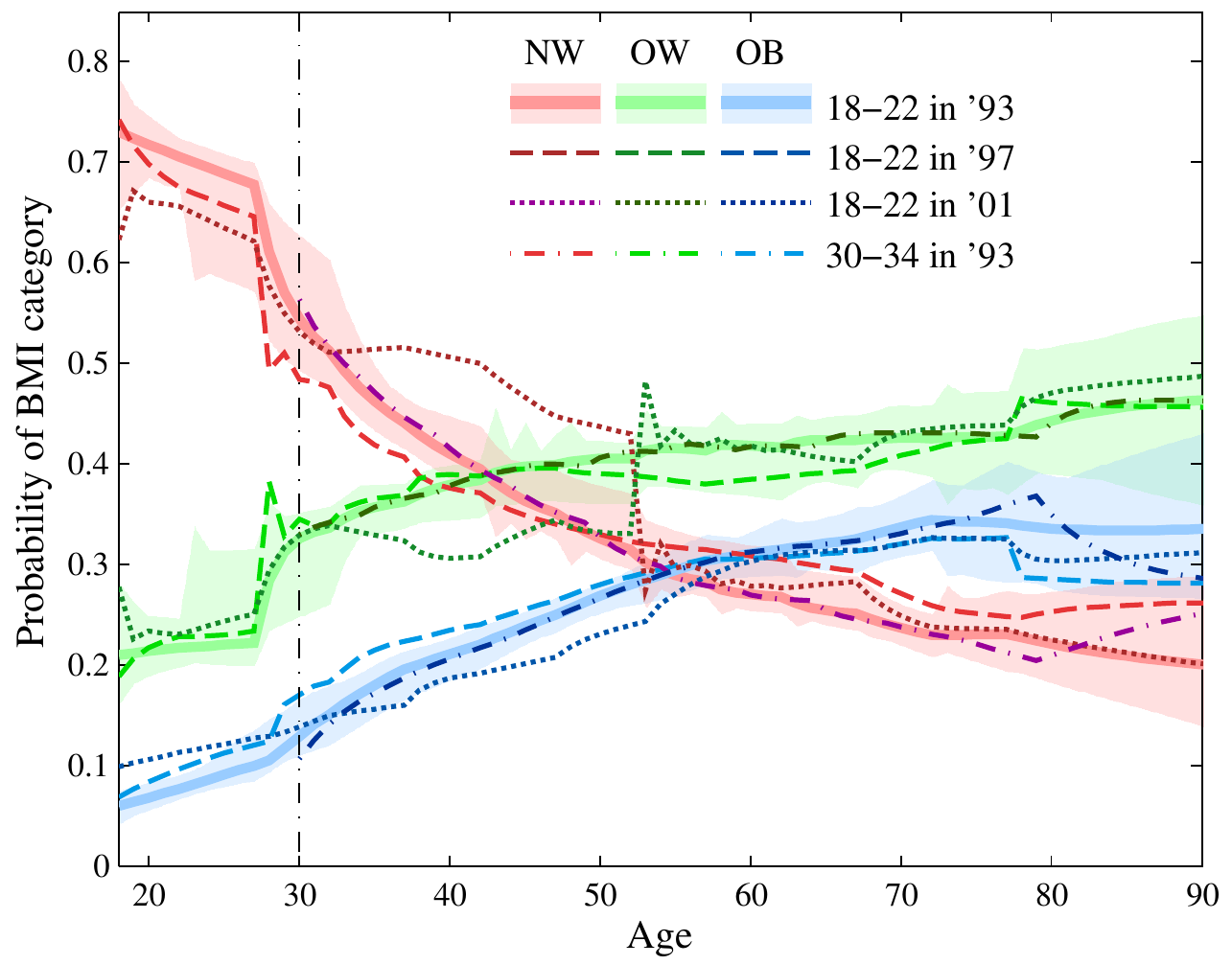}
\caption{The CSM(0)$_\text{reg}$ fit and projection of BMI trends for four cohorts of English population: aged 30--34 in 1993, aged 18--22 in 1993 (with 95\% confidence intervals), aged 18--22 in 1997 and aged 18--22 in 2001, based on the HSE data for years 1993--2013.}
\label{fig:csmcohort}
\end{figure}

We employ the CSM(0)$_\text{reg}$ model to calculate the BMI trends for birth cohorts, as described in Sec.\,\ref{sec:ageing_cohorts}. We analyse four cohorts: 30--34-years-old in 1993, 18--22-years-old in 1993, 18--22-years-old in 1997 and 18--22-years-old in 2001, throughout their adulthood. As presented in Fig.\,\ref{fig:csmcohort}, all of them display identical tendencies: the BMI distributions in the cohorts are very similar and weight gain occurs with age at the same rate.

The above CSM model results combined suggest that, contrary to the popular opinion, the excessive weight gain is experienced by the whole English population throughout the adult life, not just younger generations. This may suggest that it is driven by a common factor rather than pertains to individual lifestyles specific to certain cohorts.

\subsection{Cross-sectional data on marijuana use among US teenagers}
\label{sec:mj}

\begin{figure}[!htp]
\centering\includegraphics[width=0.69\linewidth, trim=0px 10px 0px 0px]{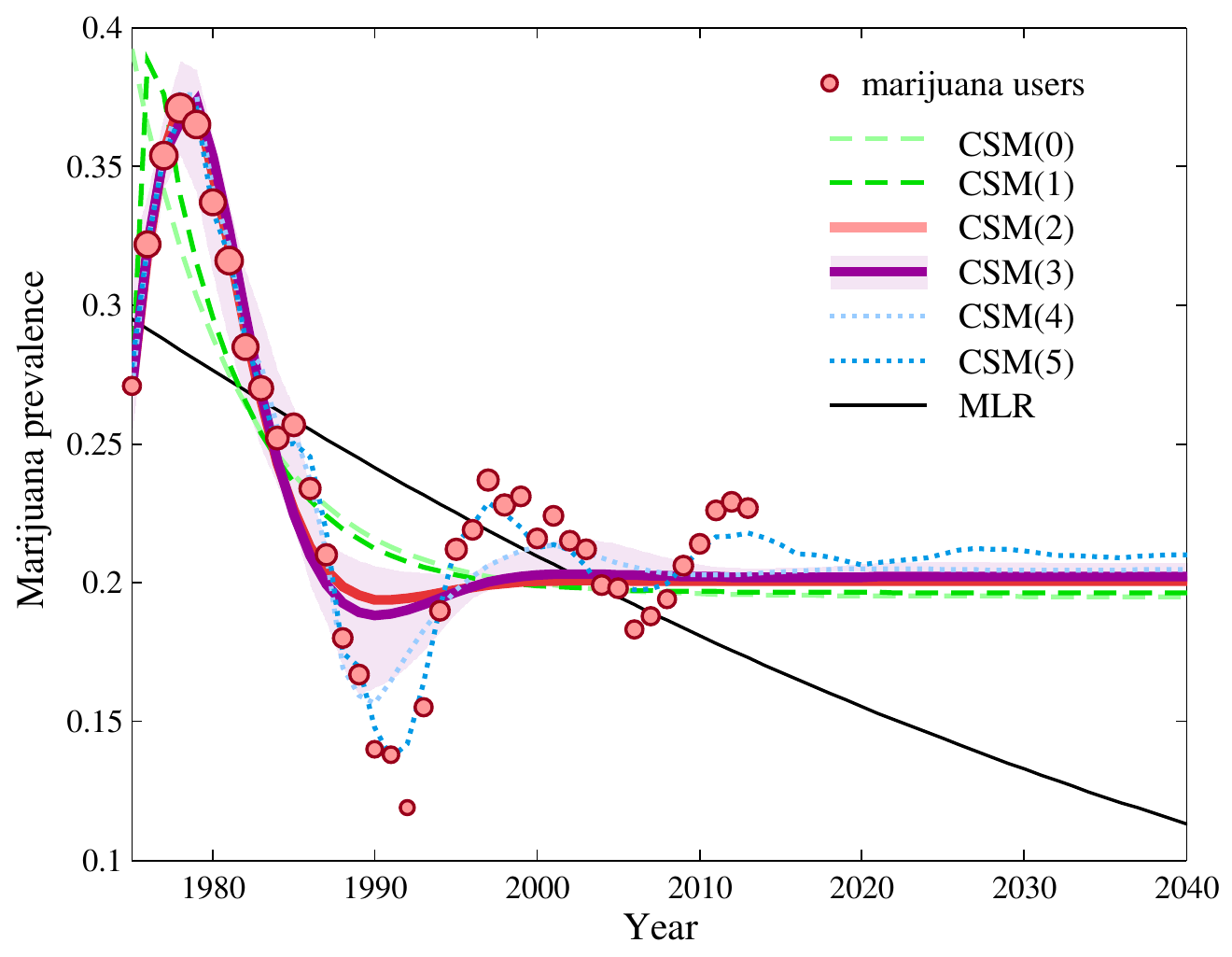}
\caption{Fits and extrapolations (with 95\% confidence interval) of marijuana use prevalence among American 12$^\text{th}$-graders based on MTF survey from years 1975--2011 obtained by CSM($\lambda$) models with different memory length of $\lambda$ years and MLR. The marker area is proportional to the observed category count.}
\label{fig:mj}
\end{figure}

The repeated cross-sectional data on the recreational use of marijuana (in the past month) among American 12$^{\text{th}}$-graders were collected from Monitoring the Future (MTF) survey in years 1975--2011 (39 independent samples of varying size, ca.~15000 on average)~\cite{marijuana}. Their trends shown in Fig.\,\ref{fig:mj} evince a complex generating mechanism steered by changing state and federal policies, inviting more sophisticated statistical analysis. In this purpose we test five memory variants of the CSM model (from memory-less to 5-year memory) and the MLR method.

Figure\,\ref{fig:mj} presents the obtained extrapolated fits of the marijuana use prevalence. All CSM models reproduce its overall long-term tendency, converging to a similar steady state of around 20\% users, while increasing the memory length enables us to recover more of its details. In contrast, the MLR method is too constrained to describe the data: it doesn't capture their rich trend and converges fast to a steady state (no drug users in the population) which is imposed by the mathematical structure of the method rather than founded on the data.

\begin{figure}[!htp]
\centering\includegraphics[width=0.36\linewidth, trim = 20 30 20 5]{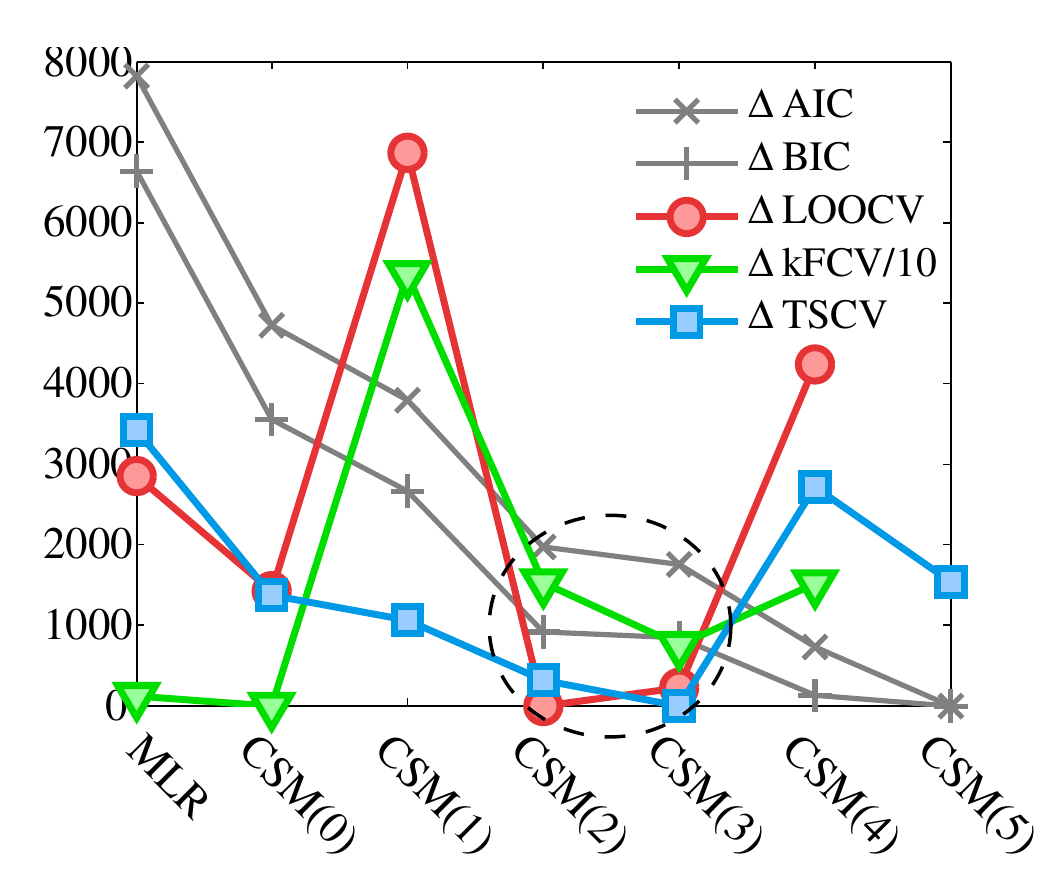}
\caption{Model selection results for the marijuana dataset, based on Table\,\ref{tab:MJerrors}. Note that $\Delta$\textit{k}FCV values were rescaled for visibility.}
\label{fig:MJ_cvplot}
\end{figure}

We verify our observations by performing the model selection procedure summarised in Table\,\ref{tab:MJerrors}. In the setting of complex data trends, AIC and BIC values are dominated by the maximum log-likelihood of the model parameters, allowing to accommodate more subtle effects. They favour CSM models with the longest memory, with more degrees of freedom than observed prevalence values. Thus, their choice should be treated with caution and a more precise and reliable model assessment provided by the cross-validation techniques is required. Accordingly, LOOCV picks CSM(2) as the best candidate, with CSM(3) relatively close to it. At the same time, $k$FCV---due to its insensitivity to the details of observed trend---selects CSM(0), but also has a local minimum for models with non-zero memory at CSM(3). The latter is additionally preferred by TSCV, with CSM(2) next to it. The cross-validation rejects CSM models with longer memory, despite a better fit to data, suggesting that it is obtained at the expense of stability of the model predictions. The MLR method is indeed insufficient to describe the analysed dataset. Figure\,\ref{fig:MJ_cvplot} helps to visualise the model selection results, identifying CSM(3) as the best model, followed by CSM(2). We conclude that although CSM(3) does not exhibit a close fit to data, it provides stable and reasonable forecasts for future applications.

\begin{table*}[!htp]
\centering
{\small\begin{tabular}{|l|c|c|c|c|c|c|c|}
\hline
Model & DOF & K-L div. & $\Delta$AIC & $\Delta$BIC & $\Delta$LOOCV & $\Delta$\textit{k}FCV & $\Delta$TSCV \\
\hline
MLR & 2 & 4139 & 7818 & 6638 & 2854 & 1200 & 3424 \\
CSM(0) & 3 & 2590 & 4723 & 3553 & 1422 & 0 & 1366 \\
CSM(1) & 7 & 2123 & 3797 & 2664 & 6866 & 53538 & 1062 \\
CSM(2) & 15 & 1203 & 1972 & 915 & 0 & 15292 & 320 \\
CSM(3) & 31 & 1077 & 1754 & 847 & 223 & 7526 & 0 \\
CSM(4) & 63 & 533 & 729 & 125 & 4243 & 15149 & 2712 \\
CSM(5) & 127 & 105 & 0 & 0 &  &  & 1533 \\
\hline
\end{tabular}}
\caption{Model selection results for the marijuana dataset. For \textit{k}FCV, $k=5$ and the averaging is performed over 300 iterations. LOOCV and $k$FCV results for the memory length $\lambda=5$ are not provided due to computational limitations (note that their values increase sharply already for $\lambda=4$).}
\label{tab:MJerrors}
\end{table*}

\subsection{Longitudinal BMI data on the US population}
\label{sec:longit-bmi}

We apply the CSM framework to longitudinal data, affected by attrition and non-adherence, on BMI collected from the National Longitudinal Survey of Youth 1979 (NLSY79) run by the U.S.~Bureau of Labor Statistics~\cite{nlsy79}. The dataset consists of 12686 trajectories belonging to men and women from a cohort born in 1957--65, interviewed in years 1981, 1982, 1985, 2006, 2008, 2010 and 2012. The BMI values were calculated from self-reported body weights and heights, and classified to one of the three categories: normal weight (NW), overweight (OW) or obese (OB) defined in Sec.\,\ref{sec:bmi}.

\begin{figure}[!htp]
\centering\includegraphics[width=0.48\linewidth, trim=10px 5px 10px 15px]{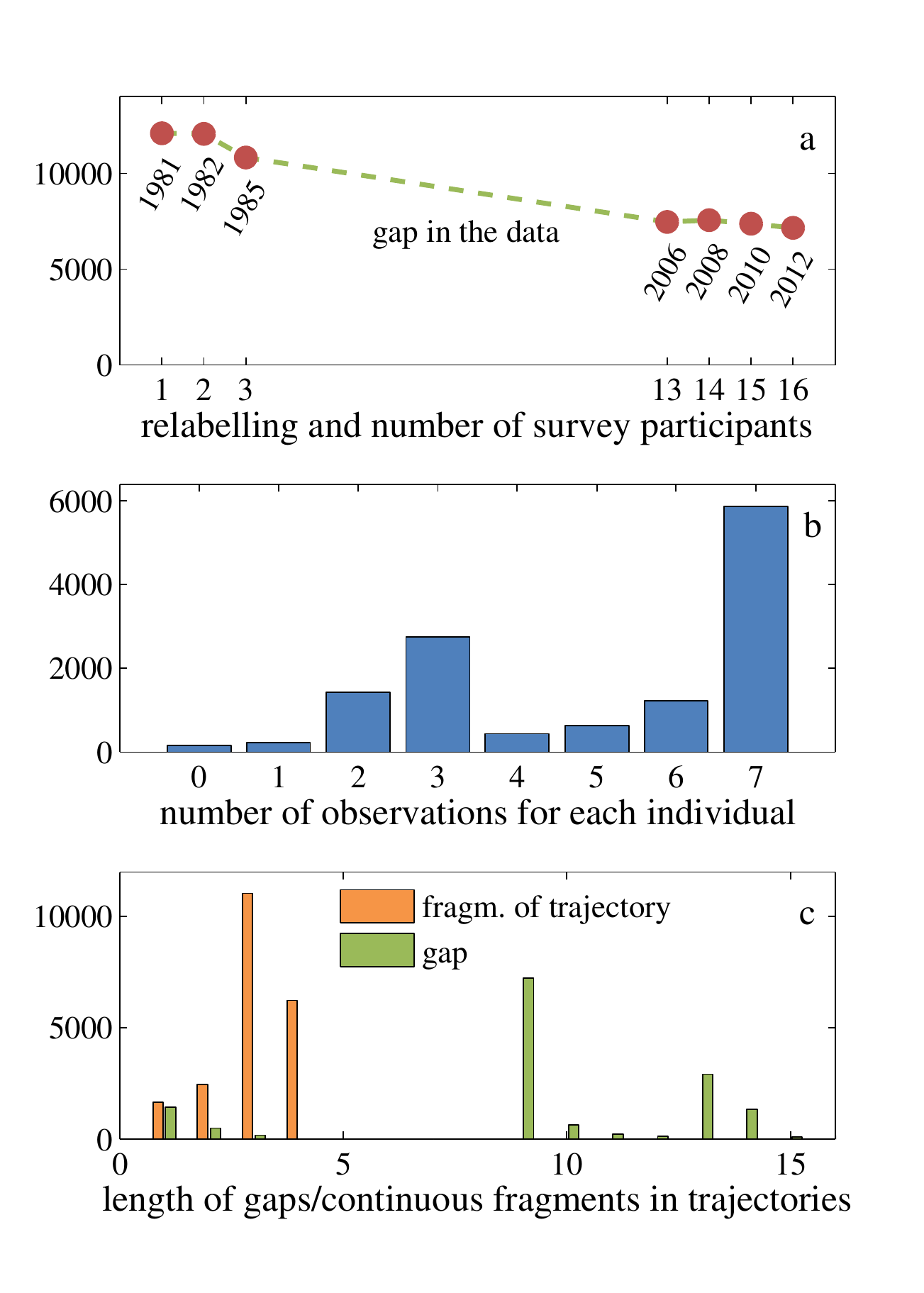}
\caption{NLSY79 data properties: a) data structure and replacing the interview years with labels (1,2,3,13,14,15,16) to improve numerical performance, one period equals approximately 2~years, b) distribution of the number of data points collected for survey participants, c) distribution of lengths of continuous fragments of data and gaps in the trajectories.}
\label{fig:histogram_csm}
\end{figure}

The properties of the investigated data are displayed in Fig.\,\ref{fig:histogram_csm}. Since the interviews were conducted at irregular time intervals, we adjust their numbering as illustrated in panel `a' to prevent numerical instabilities and speed up the calibration process, while introducing minor inaccuracies. Attrition after the first three interviews is significant, apparently due to the intermission between years 1985 and 2006 (9 intervals), leaving only 60\% of participants in the last four rounds. Yet almost all of them completed the survey with full adherence, as indicated by panel `b'. The longest continuous fragments of trajectories consist in 3 or 4 consecutive data points corresponding to the initial or final rounds of interviews, respectively, and comprise 80\% of the set of observed continuous trajectory fragments, as presented in panel `c'. The most frequent gap in the data corresponds to the survey intermission.

\begin{figure}[!htp]
\centering\includegraphics[width=0.68\linewidth, trim=0px 10px 0px 0px]{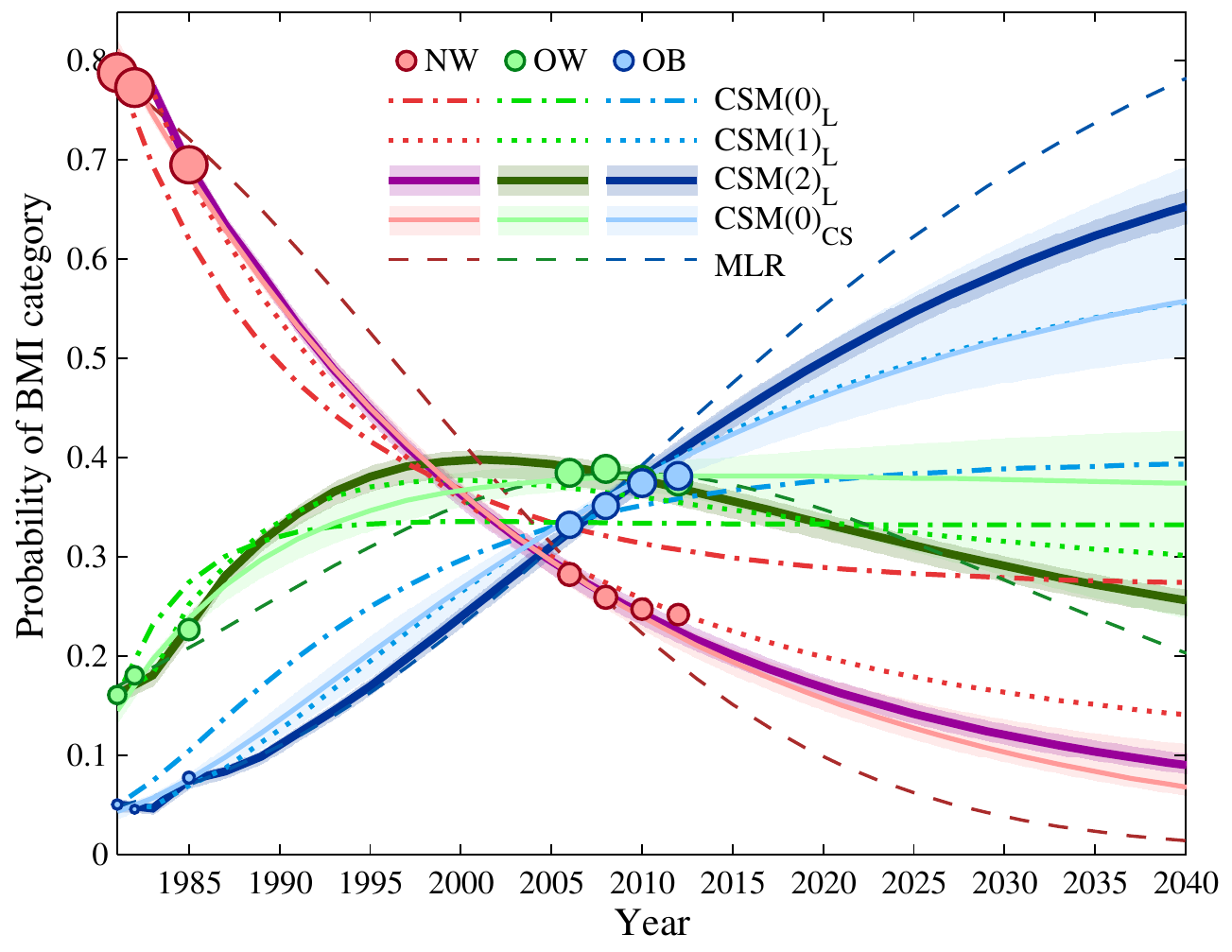}
\caption{BMI fits and predictions based on NLSY79 longitudinal survey. We compare models with different memory lengths (approx.~2$\lambda$ years, see relabelling in~Fig.\,\ref{fig:histogram_csm}): calibrated to the original, longitudinal data---CSM($\lambda$)$_\text{L}$, and calibrated to their cross-sectional reduction---CSM($\lambda$)$_\text{CS}$ and MLR.}
\label{fig:nlsy79}
\end{figure}

We calibrate different variants of the CSM model to the relabelled longitudinal data and their cross-sectional reduction defined by Eq.\,\eqref{eq:reduction}; the MLR method is used in the latter case for comparison. Figure\,\ref{fig:nlsy79} presents the extrapolated fits of the BMI distributions. According to the model selection procedure for the cross-sectional data summarised in Table\,\ref{tab:NLSBMI-cs}, BIC and TSCV choose CSM(0)$_\text{CS}$, while LOOCV and $k$FCV prefer MLR. Models with memory, for instance CSM(1)$_\text{CS}$ selected by AIC, are affected by overfitting. However, the two best models give markedly different results: CSM(0)$_\text{CS}$ achieves the steady state comprised of 37\% overweight and 63\% obese persons (through transitions between neighbouring categories, making the regularisation unnecessary), whereas MLR converges to an entirely obese population.

\begin{table}[!htp]
\centering
{\small
\begin{tabular}{|l|c|c|c|c|c|c|c|}
\hline
Model & DOF & K-L div. & $\Delta$AIC & $\Delta$BIC & $\Delta$LOOCV & $\Delta$$k$FCV & $\Delta$TSCV \\
\hline
MLR & 4 & 94.85 & 143.55 & 28.43 & 0 & 0 & 104.27 \\
CSM(0)$_\text{CS}$ & 8 & 62.16 & 86.17 & 0 & 349.02 & 346.54 & 0 \\
CSM(1)$_\text{CS}$ & 26 &1.06 & 0 & 44.09 & 5761 & 11563 & 44.05\\
\hline\hline
Model & DOF & -l.l. & $\Delta$AIC & $\Delta$BIC &  & $\Delta$$k$FCV & $\Delta$TSCV \\
\cline{1-5}\cline{7-8}
CSM(0)$_\text{L}$ & 8 & 40432 & 5231 & 4828 &  & 2452 & 2761 \\
CSM(1)$_\text{L}$ & 26 & 38317 & 1035 & 733.4 &  & 355.7 & 0 \\
CSM(2)$_\text{L}$ & 80 & 37744 & 0 & 0 & & 0 & {\color{grey}8573} \\
\hline
\end{tabular}}
\caption{Model selection results for reduced (top panel) and original (bottom panel) NLSY79 data. Since calculation of Shannon entropy is numerically difficult for longitudinal data due to the large state space for trajectories, we present the fit error as minus log-likelihood of data. $k$FCV results were calculated for 300 and 30 iterations for reduced and original data, respectively. LOOCV has been skipped as ineffective, while TSCV result for CSM(2)$_\text{L}$ is distorted by the survey intermission.}
\label{tab:NLSBMI-cs}
\end{table}

Describing the longitudinal information contained in the original data requires additional model parameters. This intuition is confirmed by the results of selection procedure carried out for models with different memory lengths, the best of which are presented in Table\,\ref{tab:NLSBMI-cs}. The simplest model CSM(0)$_\text{L}$ is unable to accommodate all available information and consequently produces a worse fit of the observed BMI trends than CSM(0)$_\text{CS}$, despite having the same number of degrees of freedom (see Fig.\,\ref{fig:nlsy79}). The best candidate is CSM(2)$_\text{L}$, yielding the most satisfactory overall model selection results (we ignore the TSCV outcome distorted for models with long memory by the survey intermission). Its estimation of the BMI distribution trends shown in Fig.\,\ref{fig:nlsy79} is similar to that obtained from CSM(0)$_\text{CS}$, which signifies the stability and effectiveness of the CSM framework. The obtained steady state consists in 3\% NW, 17\% OW and 80\% OB persons.\footnote{It is worth noting that the steady state corresponds to a stable distribution which would be achieved after a long enough time if the dynamics recovered by the CSM model from the observed data persisted unchanged in the future. Hence, the possibly shocking prediction of the obesity level in the US society may not come to pass if social customs, consumption patterns or food quality change.} Including the longitudinal information significantly narrows the confidence intervals, giving more precise estimates of analysed effects. Lastly, our results suggest that the MLR projections based on cross-sectional data may be unrealistic and prompt misleading conclusions in general.

\begin{figure}[!htp]
\centering\includegraphics[width=0.69\linewidth,trim=0pt 15pt 0pt 10pt]{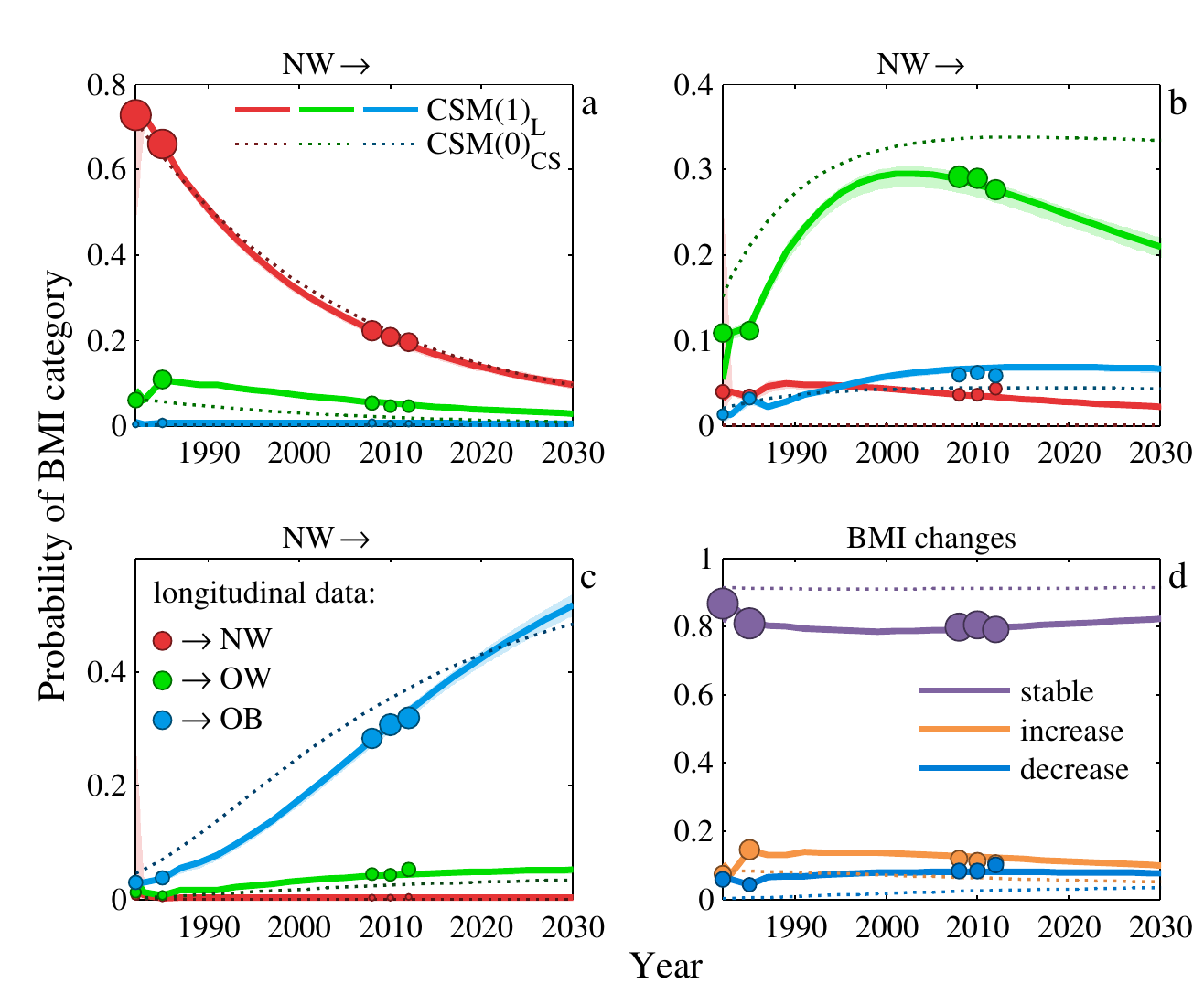}
\caption{Probabilities of moving between BMI categories in one simulation step, i.e.~2 years (with 95\% confidence interval) according to CSM(2)$_\text{L}$; CSM(0)$_\text{CS}$ and the observed longitudinal transitions (the marker area is proportional to the category count) are shown for comparison.}
\label{fig:csm_2}
\end{figure}

Another valuable information produced by the CSM framework is the temporal variability of the characteristic, encoded in the joint probabilities of belonging to particular current and past categories. We obtain it by applying Bayes' theorem\footnote{In the case of non-zero memory $\lambda$ of the process, one can simply use the extrapolated joint probability distributions $\vec{q}_t$ directly (for $\lambda =1$) or after summing over the times earlier than $t-1$ (for $\lambda > 1$).} to the transition matrix $\hat{\pi}$ and the extrapolated probability distributions $\hat{\vec{p}}_t$. Figure\,\ref{fig:csm_2} displays such joint probabilities of persons' BMI, calculated using CSM(2)$_\text{L}$. Panels `a'--`c' indicate whether they remain in the same or move to a different category within the following 2 years. The calculated trends match the values derived directly from the data by counting transitions in all available continuous two-point fragments of observed trajectories. The summary of the results is presented in panel `d': over 80\% of the population stays in the same category, while the rest tends to experience a BMI increase rather than a decrease in the short period. We also indicate respective CSM(0)$_\text{CS}$ results to demonstrate that only calibration to longitudinal data can accurately reproduce the joint probability trends.

\begin{figure}[!htp]
\centering\includegraphics[width=0.915\linewidth,trim=0pt 10pt 0pt 0pt]{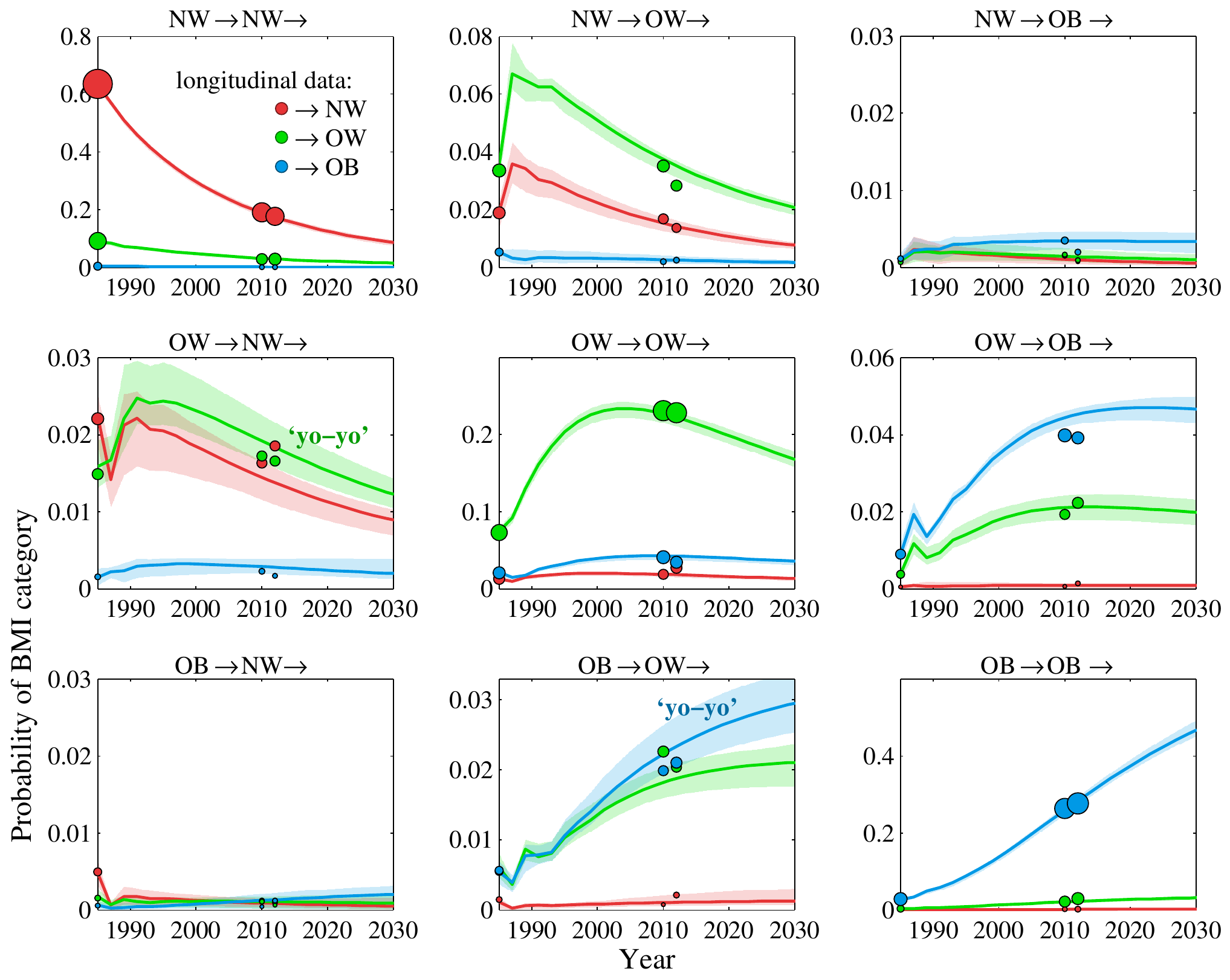}
\caption{Probabilities of moving between the BMI categories in two simulation steps, i.e.~4 years (with 95\% confidence interval) according to CSM(2)$_\text{L}$, as compared to the observed longitudinal transitions (the marker area is proportional to the category count).}
\label{fig:csm_3}
\end{figure}

Figure\,\ref{fig:csm_3} presents a similar analysis for joint probabilities of belonging to particular BMI categories at three contiguous time steps, facilitating the analysis of long-term (approx.~4 years) BMI changes using CSM(2)$_\text{L}$. We can distinguish three stable patterns: BMI remains unchanged for the majority of the population (diagonal panels) in concordance with the short term predictions; most persons who have moved to a higher BMI category remain in it (above-diagonal panels); over a half of those who have managed to reduce their BMI experience the ``yo-yo'' effect, i.e.~the cyclical loss and gain of weight (below-diagonal panels). These results are compared with the longitudinal data by counting transitions in all available continuous three-point fragments of observed trajectories. Since their number is much smaller than in the previous case (see~Fig.\,\ref{fig:histogram_csm}c), some disagreement is likely. The CSM framework unifies all available data facilitating trend analysis and forecasts of BMI tendencies in the long period.

\section{Summary}

The presented CSM framework can utilise any type of available information, from cross-sectional to longitudinal data and their mixture, for comprehensive studies of trends in groups and cohorts of the population. Its mathematical structure based on a classic Markov model has a clear dynamical interpretation, providing an insight into mechanisms generating the observed process and making it particularly well adapted to microsimulation modelling. The employed maximum likelihood estimation procedure is compatible with popular model scoring methods, while the efficient numerical implementation facilitates an extensive cross-validation of the model results and an accurate estimation of confidence intervals by bootstrapping.

The versatility of the CSM approach enables us to analyse simple dependencies, as well as complex trends, obtaining realistic projections and steady states, while avoiding ecological fallacy and overfitting. The provided examples of model applications to real-world data yield new and interesting results. In particular, the combined results on BMI trends in the English population and its birth cohorts based on cross-sectional data show that the excessive weight problem affects all generations equally, suggesting a common driving factor. The rich, shaped by historical policies, trend of marijuana use among American teenagers has been recovered assuming a 3-year-long memory of the process and shown to have achieved its steady state of about 20\% users. The result is reasonable and close to the most recent data from the same survey (21.3\% in 2015). We have described the interesting dynamics of BMI changes behind the obesity growth in the US population concealed in incomplete longitudinal data (e.g.~`yo-yo' effect), which cannot be recovered from the longitudinal information alone. These findings can be particularly useful when designing interventions to prevent the shocking scenario of the steady state 80\% obesity level found by the model. The above data analyses have included the model selection procedure, which enabled us to choose the most appropriate variants of the CSM model and revealed that the commonly used logistic regression can produce incorrect and misleading results.

\appendix

\section{Error estimation by bootstrapping}
\label{sec:bootstrapping}

Bootstrapping is a very general method of calculating confidence intervals for quantities estimated from statistical data~\cite{Efron1993}. In its simplest form, we construct an empirical distribution of such a quantity by drawing randomly with replacement from the dataset to obtain a new sample, from which a new value of the quantity can be computed. By doing it a sufficient number of times we can calculate e.g.~95\% confidence intervals for the quantity. A more advanced version, parametric bootstrapping, first estimates the distribution of the data based on the observed sample (e.g.~using Bayesian inference) and then draws from this distribution.

In our case, the datasets consist in sets of surveys collected at multiple times or observed longitudinal trajectories. Our quantity can be the transition matrix $\fit\mpi$, initial state $\fitvp_0$ or extrapolated distribution $\fitvp_t$ inferred by the CSM model. In the case of cross-sectional data, for each time $t$ with a non-zero number $\obs{n}_t$ of surveys, we assume a flat Dirichlet prior of $\vec{p}_t$, and thus its posterior distribution is $\mathcal{D}_{\vec{\alpha}_t}$ with $\alpha_{t,k} = 1 + \obs{n}_{kt}$. We draw from $\mathcal{D}_{\vec{\alpha}_t}$ a new distribution $\vec{p}'_t$ and next draw from it a set of $\obs{n}_t$ new values of $X_t$. A resampled distribution $\obsvp'_t$ is obtained by counting $\obs{n}'_{kt}$ occurrences of each $k$. For longitudinal data, given the set of observed trajectories, we draw with replacement a new set of the same size. We use this simple procedure instead of estimating a Dirichlet distribution because its dimension can be very large in this case. The above approach respects the observed trends in the data and is free from assumptions about the error distribution or analytical simplifications (unlike the commonly used ``delta method'' mentioned in Sec.\,\ref{sec:bayesian-analysis}).

Having generated a new input set $\obsvp'_t$, we obtain a new transition matrix $\fit\mpi'$, initial state $\fitvp'_0$, and consequently a new extrapolated trend $\fitvp'_t$ following Sec.\,\ref{sec:mle}. To obtain $\alpha$ confidence intervals for each $\fitp_{kt}$, we sort the bootstrapped trends $\fitvp'_t$ by their total Kullback--Leibler divergence from $\fitvp_t$, $\sum_{t=0}^{T_2-1} \DKL(\fitvp_t\Vert\fitvp'_t)$, where $T_2 - T$ is the number of extrapolated periods, and remove the furthest $1-\alpha$ trends. The upper and lower confidence bounds for $\fitp_{kt}$ are the maximum and minimum of the remaining $\fitp'_{kt}$ values. This also provides the confidence intervals for $\fitvp_0$. A similar algorithm can be applied to $\fit\mpi$ by treating its columns as probability distributions and sorting the bootstrapped matrices $\fit\mpi'$ by their total Kullback--Leibler divergence from $\fit\mpi$ (summing over columns).

\end{document}